\documentclass[numberedappendix]{emulateapj}
\usepackage{epsfig}
\usepackage{graphicx}
\usepackage{color}
\usepackage{enumerate}

\newcommand\kms{{\rm km~s^{-1}}}

\def\msun{\,{\rm M_\odot}}
\newcommand\rsun{{\rm R_\odot}}

\def\ndot{\,{\rm yr^{-1}}}
\def\E{{E}}
\def\J{{J}}

\begin{document}

\shorttitle{Tidal stellar disruptions by massive black hole pairs}
\shortauthors{Chen et al.}
\slugcomment{}

\title{Tidal stellar disruptions by massive black hole pairs: II. Decaying binaries}

\author{Xian Chen\altaffilmark{1,2,5}, Alberto Sesana\altaffilmark{3,4}, Piero Madau\altaffilmark{2}, \& F. K. Liu\altaffilmark{1,5}}

\altaffiltext{1}{Department of Astronomy, Peking University, 100871 Beijing, China; {\it fkliu@pku.edu.cn}}
\altaffiltext{2}{Department of Astronomy \& Astrophysics, University of 
California, 1156 High Street, Santa Cruz, CA 95064.}
\altaffiltext{3}{Center for Gravitational Wave Physics, The Pennsylvania State University, University 
Park, State College, PA 16802.}
\altaffiltext{4}{Max Planck Institute for Gravitationalphysik (Albert Einstein Institute), Am M\"uhlenberg , 14476, Golm, Germany}
\altaffiltext{5}{Kavli Institute for Astronomy and Astrophysics, Peking University, 100871 Beijing, China.}

\begin{abstract}
Tidal stellar disruptions have traditionally been discussed as a probe of the single, massive black holes (MBHs) that are dormant 
in the nuclei of galaxies. In Chen et al. (2009), we used numerical scattering experiments to show that three-body interactions 
between bound stars in a stellar cusp and a non-evolving ``hard" MBH binary will also produce a burst of tidal disruptions, caused 
by a combination of the secular ``Kozai effect" and by close resonant encounters with the secondary hole. Here we derive basic analytical scalings of the stellar disruption rates with the system parameters,
assess the relative importance of the Kozai and resonant encounter mechanisms as a function of time,
discuss the impact of general relativistic (GR) and  extended stellar cusp effects, and develop a hybrid 
model to self-consistently follow the shrinking of an MBH binary in a stellar background, including slingshot 
ejections and tidal disruptions.
In the case of a fiducial binary with primary hole mass $M_1=10^7\,\msun$ and mass ratio $q=M_2/M_1=1/81$, embedded 
in an isothermal cusp, we derive a stellar disruption rate $\dot N_* \sim 0.2\,$yr$^{-1}$ lasting $\sim 3\times 10^5$ yr. 
This rate is 3 orders of magnitude larger than the corresponding value for a single MBH fed by two-body relaxation, confirming
our previous findings. For $q\ll0.01$, the Kozai/chaotic effect could be quenched due to GR/cusp effects by an order of magnitude, but even in this case the stellar-disruption rate is still two orders of magnitude larger than that given by standard relaxation processes around a single MBH. Our results suggest that $\ga10\%$ of the tidal-disruption events may originate in MBH binaries.
\end{abstract}


\section{INTRODUCTION}\label{intro}

Stars that wander too close to the MBHs that reside at the center of galaxies are shredded by the tidal 
gravitational field of the hole. After a tidal disruption event, about half of the debris are spewed into eccentric bound orbits and
fall back onto the hole, giving rise to a bright UV/X-ray outburst that may last for a few years \citep[e.g.][]{rees88}. 
``Tidal flares" from MBHs may have been observed in several nearby inactive galaxies \citep{komossa02,esquej07}. The 
inferred stellar disruption frequency is $\sim10^{-5}~{\rm yr^{-1}}$ per galaxy (with an order of magnitude 
uncertainty, \citealt{donley02}), comparable to the theoretical expectations for single MBHs fed by two-body relaxation \citep{wang04}.

Yet, MBHs are not expected to grow in isolation. According to the standard paradigm of structure 
formation in the universe, galaxies merge frequently during the assembly of their dark matter halos. As MBHs
become incorporated into larger and larger halos, they sink to the center
of the more massive progenitor owing to dynamical friction from distant stars,
and form bound binaries (MBHBs). In a purely stellar background, as the binary separation
decays, the effectiveness of dynamical friction slowly declines, and the pair then ``hardens" via
three-body interactions, i.e. by capturing stars that pass close to the
holes and ejecting them at much higher velocities \citep[e.g.][]{begelman80,quinlan96,volonteri03,sesana06}. If the hardening continues sufficiently far, 
possibly driven by efficient stellar relaxation processes in a triaxial potential \citep[e.g.][]{merritt04} or in the presence of 
massive perturbers \citep[e.g.][]{perets07,perets08}, or by dissipative gaseous processes \citep[e.g.][]{colpi09},
gravitational radiation losses finally take over, and the two MBHs will coalesce in less than a Hubble time \citep[e.g.][]{mm05,sesana05,sesana07}.
In \citet{chen09}, we used scattering experiments to show that gravitational slingshot interactions between a non-evolving, unequal-mass 
hard binary and a bound stellar cusp will inevitably be accompanied by a burst of stellar tidal disruptions. Our work differed 
from those by \citet{ivanov05}, who developed an analytical theory of the secular evolution of stellar orbits in the gravitational
field of a MBHB, and by \citet{chen08}, who argued that stellar disruption rates by MBHBs fed by two-body relaxation would be smaller than
those expected for single MBHs. Our numerical experiments revealed that a significant fraction of stars initially 
bound to the primary hole are scattered into its tidal disruption loss cone by resonant interactions with the secondary 
hole, close encounters that change the stellar orbital parameters in a chaotic way. 

In this paper we continue our investigations of stellar disruptions by MBHBs embedded in bound stellar cusps. We develop a hybrid 
model that self-consistently follows over time the shrinking of an MBH binary, the evolution of the stellar cusp, and the stellar disruption rate.
The plan is as follows. In \S~\ref{background}, we introduce the basic theory of stellar disruption processes by MBHB systems. We describe 
our numerical scattering experiments in \S~\ref{experiments}, and discuss our results for different binary parameters as well as the 
effect of general relativistic corrections in \S~\ref{tests}. A detailed study of the properties of disrupted stars is carried 
out in \S~\ref{properties}.  As a first step towards understanding the dependence of stellar consumptions on the parameters of the system, 
in \S~\ref{lcfr} we fix the binary semimajor axis and its eccentricity, and calculate the stellar disruption rate 
in the stationary case. In \S~\ref{sdr}, we present our hybrid model and calculate the disruption rates for an evolving, shrinking MBHB. 
Finally, we summarize and discuss our results in \S~\ref{dc}.

\section{BASIC THEORY OF STELLAR DISRUPTIONS}\label{background}

Consider an isotropic background of stars all of mass $m_*$ and radius $r_*$, 
centered on an MBH. Let $\Psi(r)$ be the total gravitational potential at radius $r$, and $r_t=r_*(M_{\rm BH}/m_*)^{1/3}$ 
the tidal disruption radius, 
\begin{equation}\label{rt}
r_t\simeq 5\times10^{-6}~{\rm pc}~\left(\frac{r_*}{\rsun}\right)\left(\frac{\msun}{m_*}\right)^{1/3}
\left(\frac{M_{\rm BH}}{\rm 10^7\,\msun}\right)^{1/3}.
\end{equation}
The phase-space region of specific energy $E_*$ and specific angular momentum $J_*$ bounded by
\begin{equation}\label{jlcrt}
\J_{\rm lc}^2(\E_*,r_t)=2r_t^2[\E_*-\Psi(r_t)]
\end{equation}
is populated by stars on orbits crossing $r_t$, and thus susceptible to tidal disruption. We name this cone-like region of phase space 
the ``tidal loss cone". Whether the tidal loss cone can be emptied by stellar disruption depends on the efficiency of stellar relaxation. 
Let $T_r(\E_*)$ be the relaxation timescale of stars with specific energy $\E_*$, $P_*(\E_*)$ their orbital period, and $\J_c(\E_*)$ the 
specific angular momentum of a circular orbit with energy $\E_*$. In the ``pinhole limit'' \citep{lightman77}, $P_*(\E_*)/T_r(\E_*)\gg 
\J_{\rm lc}^2(\E_*,r_t)/\J^2_c(\E_*)$, a star can random walk in and out of the tidal loss cone within one orbital period, and the tidal loss 
cone remains almost full despite tidal disruptions. In the ``diffusion limit'', $P_*(\E_*)/T_r(\E_*)\la \J_{\rm lc}^2(\E_*,r_t)/
\J^2_c(\E_*)$, the tidal loss cone is emptied after a single orbital period, and stars diffuse into the loss cone on the 
relaxation timescale. 
Assume now that the central primary hole of mass $M_1$ forms a binary pair with a secondary hole of mass $M_2<M_1$, and let $a$ be the 
semimajor axis of the system. The $\E_*-\J_*$ region of phase space bounded by
\begin{equation}\label{intjlc}
\J_{\rm lc}^2(\E_*,a)=2a^2[\E_*-\Psi(a)]
\end{equation}
is composed of orbits that are either inside or intersect a sphere of radius $a$. If the binary is ``hard", a star on such orbit will 
undergo a three-body interaction with the MBHB, so we refer to the phase space defined by equation~(\ref{intjlc}) as the ``interaction 
loss cone''. Three-body interactions perturb the energy and angular momentum of ``intruder" stars, acting as an additional source 
of stellar relaxation. If three-body relaxation occurs in the diffusion regime, the stellar consumption rate will be enhanced.

To proceed further, we must first define some characteristic scales of a MBHB system. Recent numerical simulations have shown that 
three-body interactions between the binary and intruder stars result in significant energy exchange when the total stellar mass within the 
binary orbit is comparable to or smaller than the mass of the secondary hole \citep{baumgardt06,matsubayashi07}. We denote with $a_0$ such a 
critical binary separation: the binary shrinks by dynamical friction when $a\ga a_0$, and by three-body processes at smaller separations. 
Following \citet{sesana08}, we assume that the stellar distribution follows a double power-law with break radius $r_0$, defined 
as the radius of the ``sphere of influence" containing a mass in stars equal to $2M_1$. For $r>r_0$, the stellar density 
profile follows an isothermal distribution,
\begin{equation}\label{rhoiso}
\rho_*(r)=\frac{\sigma_*^2}{2\pi Gr^2},
\end{equation}
where $\sigma_*$ is the 1-D velocity dispersion, while for $r<r_0$ $\rho_*(r)\propto r^{-\gamma}$. 
It is easy to derive then
\begin{equation}
r_0=(3-\gamma)GM_1/\sigma_*^2\simeq4.6{~\rm pc}~(3-\gamma)M_7\sigma_{100}^{-2}
\end{equation}
and $a_0=q^{1/(3-\gamma)}r_0$, where $M_7\equiv M_1/10^7\,\msun$, $q\equiv M_2/M_1$ is the binary mass ratio, and 
$\sigma_{100}\equiv \sigma_*/100\,\kms$. Notice that the ``three-body radius" $a_0$ is larger than the conventional ``hardening" radius 
$a_h=GM_2/(4\sigma_*^2)$ \citep{quinlan96}. The ratio between the tidal radius of the primary hole, $r_{t1}$, and $a_0$,
\begin{equation}
\frac{r_{t1}}{a_0}\simeq {10^{-6} \over 3-\gamma}q^{-p}\,{\sigma_{100}^2\over 
M_7^{2/3}}\left(\frac{r_*}{\rsun}\right)
\left(\frac{\msun}{m_*}\right)^{1/3}, \label{ratio}
\end{equation}
where $p\equiv 1/(3-\gamma)$, indicates that the interaction loss cone of a binary is much larger than the tidal loss cone of a single 
MBH. Therefore, the transfer of only a small fraction of interacting stars into the tidal loss cone will cause a large enhancement of 
the stellar disruption rate.

When does the presence of a binary begin affecting the stellar disruption rate? Let us assume that, before the intrusion of 
the secondary hole, stellar relaxation is dominated by two-body interactions. 
Stars in the diffusion limit are bound to the primary hole, and their specific energy $\E_*$ is related to the orbital semimajor axis $a_*$ by 
$\E_*=-GM_1/(2a_*)$. The boundary between the pinhole and diffusion limits is then dictated by the condition
\begin{equation}\label{boundary}
{P_*(a_*)/T_r(a_*)}={\J_{\rm lc}^2(a_*,r_{t1})/\J^2_c(a_*)}={r_{t1}/a_*}.
\end{equation} 
Substituting into the above equation the two-body relaxation timescale,
\begin{eqnarray}
T_r(r)&=&\frac{\sqrt{2}\sigma_*^3}{\pi G^2m_*\rho_*(r)\ln\Lambda} \nonumber\\
&=& 5 ~{\rm Gyr}~\sigma_{100}\left(10\over \ln\Lambda\right)\left(r_0\over 1~{\rm pc}\right)^2\left(r\over r_0\right)^\gamma \label{tr}
\end{eqnarray}
(where $\ln \Lambda$ is the Coulomb logarithm) and the Keplerian orbital period of the star orbiting $M_1$, $P_*(a_*)=2\pi[a_*^3/(GM_1)]^{1/2}$,
and assuming $r_*=\rsun$ and $m_*=\msun$, we can write the critical radius $a_c$ marking the boundary between pinhole and 
diffusion regimes as
\begin{equation}\label{acri}
{a_c\over r_0}\simeq 0.33^{2/s}(3-\gamma)^{-1/s}M_7^{2/3s}\sigma_{100}^{4/s}\left(\ln\Lambda\over10\right)^{-2/s}, 
\end{equation}
where $s\equiv 5-2\gamma$. If a secondary hole is now added to the system, and the interaction loss cone is not empty, a significant 
enhancement of stellar disruptions occurs when the binary separation shrinks to $a\sim a_c$. For an isothermal density profile and 
a primary hole satisfying the $M_{\rm BH}-\sigma_*$ relation, $M_7=\sigma_{100}^4$ \citep{tremaine02}, equation (\ref{acri}) implies 
$a_c>a_0$ as long as $q<0.1\,M_7^{5/3}$, i.e. for unequal-mass binaries the enhancement of stellar disruptions starts during the 
dynamical friction early phases of the binary orbital evolution. N-body simulations have shown, however, that the secondary hole decays 
from $a_c$ to $a_0$ and enters the three-body interaction regime on a timescale $<10^5$ yr. Here, we ignore the early dynamical friction 
phases and focus on stellar disruptions induced by three-body scattering events. At binary separation $a=a_0$, the interaction 
loss cone contains stars that can be bound or unbound to the primary. A bound star can interact with the binary multiple times 
before leaving the system, significantly increasing its probability of being tidally disrupted. For equal-mass binaries, 
the radius of influence $r_0$ is comparable to the three-body radius $a_0$, and most scattering events involve stars that 
are unbound (or marginally bound). The impact 
of bound stars is more important for unequal-mass MBHBs, and these systems will be the main focus of this paper.

A bound star with semimajor axis $a_*<a/2$ never crosses the orbit of the secondary hole and undergoes a secular evolution 
in which its orbital eccentricity is excited and oscillates periodically, the so-called ``Kozai effect'' 
\citep{kozai62,lidov62,ivanov05,gualandris09}. The period of oscillation (``Kozai timescale'') is 
\begin{equation}\label{tk}
T_K(a_*)=\frac{2}{3\pi q}\left(\frac{a_*}{a}\right)^{-3/2}P(a),
\end{equation}
\citep{innanen97,kiseleva98}, where
\begin{eqnarray}\label{period}
  P(a)&=&2\pi a^{3/2} \left[G(M_1+M_2)\right]^{-1/2} \nonumber \\ 
&\simeq & 10^3~{\rm yr}~(1+q)^{-1/2} M_7^{-1/2}\left(a\over 0.1~{\rm pc}\right)^{3/2}
\end{eqnarray}
is the orbital period of the binary. Since $T_K(a)\ll T_r(a)$, the Kozai mechanism is much more efficient than two-body interactions 
at repopulating the tidal loss cone. However, when $q\ll1$, $r_0\gg a_0$ and the majority of bound stars have close encounters 
with the secondary hole that change the orbital elements of the star in a complicated chaotic way. In this regime: 1) numerical simulations 
are needed to give reasonable estimates of the tidal disruption rates; and 2) the contribution to the gravitational potential by background stars 
as well as stellar collisions can be neglected during the interaction. When $a_*\ll a$, two-body relaxation can be more efficient than Kozai 
precession in changing stellar orbits (compare eqs. \ref{tr} and \ref{tk}, and notice that $\sigma_*^2\propto a_*^{-1}$ at $a_*\ll r_0$), but the number 
of these stars is negligible. Under these conditions, the problem can be tackled by means of restricted three-body scattering experiments. 

\section{SCATTERING EXPERIMENTS}\label{experiments}

The integration of the three-body encounter equations is performed in a coordinate system
centered at the location of $M_1$. Initially, the binary (of mass ratio $q$ and eccentricity $e$) has
a randomly-oriented orbit with $M_2$ at its pericenter: stars move in the $x-y$ plane
with pericenters along the positive $x$-axis and random orbital phases. The initial conditions of the 
{\it restricted} three-body problem problem are then completely defined by 6 variables, 3 for the
binary and 3 for the star: 
1) the inclination of the orbit of the binary, $\theta$, i.e. the angle between the angular momentum of the binary and the $z$ axis; 
2) the longitude of the secondary hole ascending node, $l$;
3) the argument of the pericenter of the secondary hole, $\phi$ (if $e\neq0$);
4) the semimajor axis of the stellar orbit, $a_*$; 
5) the normalized (by the angular momentum of a circular orbit with the same semimajor axis) angular momentum of the star, $j_*$; and
6) the orbital phase of the star, $p_*$.
We start each scattering experiment by generating $6$ random numbers, with
$\cos\theta$ evenly sampled in the range $[-1,1]$, and both $l$ and $\phi$ uniformly distributed in the range
$[0,2\pi]$. We sample $a_*$ logarithmically around $a$ (the range is described in detail below) and $j_*^2$ randomly 
between 0 and 1 (corresponding to an isotropic distribution). Given the $j_*$ of a star, we numerical integrate one revolution of 
a Keplerian orbit with eccentricity $e_*=(1-j^2_*)^{1/2}$, and derive $p_*(t)$ as a function of time $t$. Then the initial 
orbital phase for the scattering experiment is drawn from the distribution function $f(p_*)=dt/dp_*$.

Having defined the initial conditions, the orbit of each star was followed by integrating the coupled first-order differential equations
\begin{eqnarray}
  \dot{\mathbf{r}} &=& \mathbf{v} \\
  \dot{\mathbf{v}} &=&
  -G\sum_{i=1}^{2}\frac{M_{i}(\mathbf{r}-\mathbf{r}_{i})}{|\mathbf{r}-\mathbf{r}_{i}|^3},\label{accel}
\end{eqnarray}
where $\mathbf{r}$ and $\mathbf{v}$ are the position and velocity vectors of the star and $\mathbf{r}_i$ is the position of the $i$th ($i=1,2$) MBH. 
When $e\neq0$, we included a subroutine to numerically compute the positions of the two holes at each timestep. The units in the numerical 
computation were $G=M_{12}=a=1$ (with $M_{12}\equiv M_1+M_2$) and the integrator was an explicit
Runge-Kutta method of order 8 \citep[\textbf{dopri8},][]{hairer87}, with a fractional error per step in position and velocity
set to $10^{-13}$. The integration was stopped if one of
the following conditions was satisfied: (1) the star left the sphere of radius $a(10^{10}q)^{1/4}$, where the quadrupole force
from the binary is ten orders of magnitude smaller than $GM_{12}/a^2$, with positive energy;
the physical integration timescale exceeded $10^{10}$ yr; (3) the number of required integration timesteps reached $10^8$. 
Conditions (2) and (3) were adopted to save computational time, as a small fraction ($\la3\%$ depending on $q$ and $e$) of stars 
are scattered into wide, bound orbits and may survive many revolutions. We have tested our code by reproducing Figures 
4 and 6 of \citet{sesana08} (who used full three-body scattering experiments), and found excellent agreement.

\section{TESTS}\label{tests}

To understand the dependence of our results on various properties of the MBHB, such as $q$ and $e$, and the impact of general 
relativistic effects, we have performed a number of tests with $N=10^4$ stars in each run. The initial semimajor axis of the intruder star 
was sampled logarithmically in the interval $[1/2a,2a]$, where three-body interactions are expected to be the strongest. 
We recorded the minimum separation between the stars and the holes during each scattering experiment, and analyzed the results 
in terms of the fraction of stars reaching a given distance from a member of the pair. In the case of unbound stars, if the initial 
distribution of pericenter distances is uniform, such fraction has the physical meaning of a close-encounter cross section \citep{chen08}. 
While for the bound stars considered here, the concept of cross section no longer strictly applies because the initial pericenter-distance distribution 
is not uniform, for convenience we shall still refer to this fraction as the close-encounter cross section in the following.

\subsection{Close-encounter cross section}

\begin{figure}
\plotone{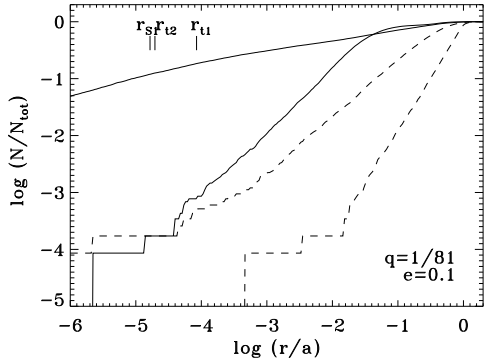}
\caption{Close-encounter probability for bound stars interacting with a MBHB of mass ratio $q=1/81$ and orbital eccentricity $e=0.1$. 
The vertical axis shows the fraction of stars $N/N_{\rm tot}$ with closest approach distance $r_{\rm min}<r$.
{\it Solid lines:} multi-encounter cross section. {\it Dashed lines:} single encounter cross-section. 
The upper solid and dashed curves refer to the primary hole, the lower ones to the secondary. The short vertical lines mark
the positions of the tidal and Schwarzschild radii of the two holes for $M_7=1$ , $\sigma_{100}=1$, $a=a_0$, and $\gamma=2$. 
The sharp cutoff at $N/N_{\rm tot}\simeq 10^{-4}$ is an artifact of small number statistics in the scattering experiments. 
}
\label{cs81e1}
\vspace{0.3cm}
\end{figure}

We performed scattering experiments for $q=1/81$ and $e=0.1$, where each star was allowed to encounter the binary as many times as required
before the integration was stopped. Then the minimum separation between the star and each hole during the entire course of the interaction
was recorded for the calculation of the ``multi-encounter cross section". We also recorded the first minimum separation (a local minimum in the 
distance-time curve) during the
encounter between the star and each hole, and calculated the ``single-encounter cross section". The latter can be viewed as the interaction probability
in the case of an isolated MBH. The resulting cross sections are plotted in Figure~\ref{cs81e1}. As already shown by \citet{chen09}, 
the multi-encounter cross section for $M_1$ is dramatically higher than the single-encounter probability: when $r\simeq 10^{-4}a$, 
corresponding to the tidal radius of a primary hole with $M_7=1$ embedded in an isothermal cusp with $\sigma_{100}=1$, at separation $a=a_0$, the 
multi-encounter cross section of $M_1$ is nearly 3 dex larger than that for single encounters. In the following, we shall refer to 
the multi-encounter cross section at $r_{t1}$ as the ``tidal disruption cross section". Because this is much larger than the corresponding
cross section for the secondary, stellar disruptions by an unequal-mass binary will be dominated by the primary hole, and will be the focus of
our analysis.

\subsection{Dependence on q and e}\label{effqe}

\begin{figure}
\plotone{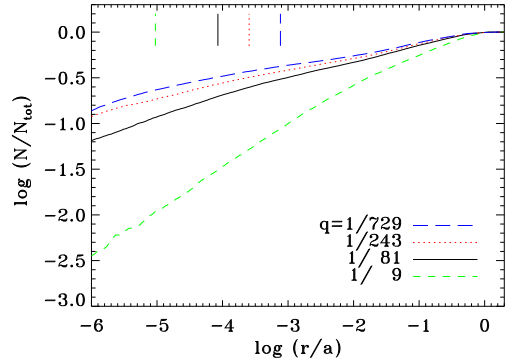}
\caption{Multi-encounter cross section for bound stars interacting with the primary hole of a MBHB with $e=0.1$ and different mass ratios $q$. 
The short vertical lines mark the locations of the tidal radii of $M_1$ for the same parameters used in Fig.~\ref{cs81e1}.
}\label{multiq}
\end{figure}

\begin{figure}
\plotone{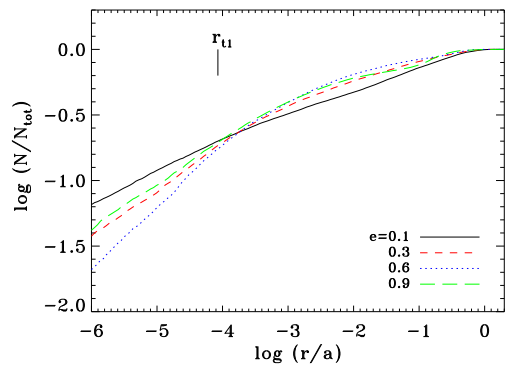}\\
\caption{Same as Fig.~\ref{multiq}, but for $q=1/81$ and different values of the binary eccentricity $e$.}\label{multie}
\end{figure}

To study the dependence of the close-encounter cross section on the binary mass ratio, we performed three additional sets of scattering experiments 
for $e=0.1$ and $q=1/9$, $1/243$, and $1/729$, each using $10^4$ stars. The results show (Figure~\ref{multiq}) that, as $q$ increases, the 
multi-encounter probability decreases from 41\% ($q=1/729$) to 1.1\% ($q=1/9$). This is because, as the perturbing force from the secondary hole 
becomes stronger, a star is more susceptible to ejection. Figure~\ref{multie} shows the dependence of the multi-encounter cross section on binary 
eccentricity at fixed $q=1/81$. For $r/a>10^{-6}$, the cross section varies at most by a factor of 3 as $e$ increases from 0.1 to 0.9. When 
$r_{t1}/a\simeq10^{-4}$, the tidal disruption probability does not depend significantly on eccentricity. 

\section{PROPERTIES OF DISRUPTED STARS}\label{properties}

To understand the physical processes responsible for the enhancement of the tidal disruption probability, we need to investigate 
the properties of the disrupted stars. We performed new scattering experiments aimed at covering the whole parameter space of the interacting stellar
population, extending the range of semimajor axis $a_*$ from $[a/2,2a]$ to $[a/20,20a]$. 
We ran four sets of numerical experiments, each consisting of $5\times 10^4$ stars, for varying binary eccentricities and $q=1/81$. A star is
counted as disrupted if its separation from the primary hole becomes smaller than $r_{t1}$. (To calculate $r_{t1}/a_0$, the fiducial 
parameters $M_7=1$, $\sigma_{100}=1$, and $\gamma=2$ were used.) 

\subsection{Phase-space distribution}\label{phasespace}

\begin{figure*}
\plottwo{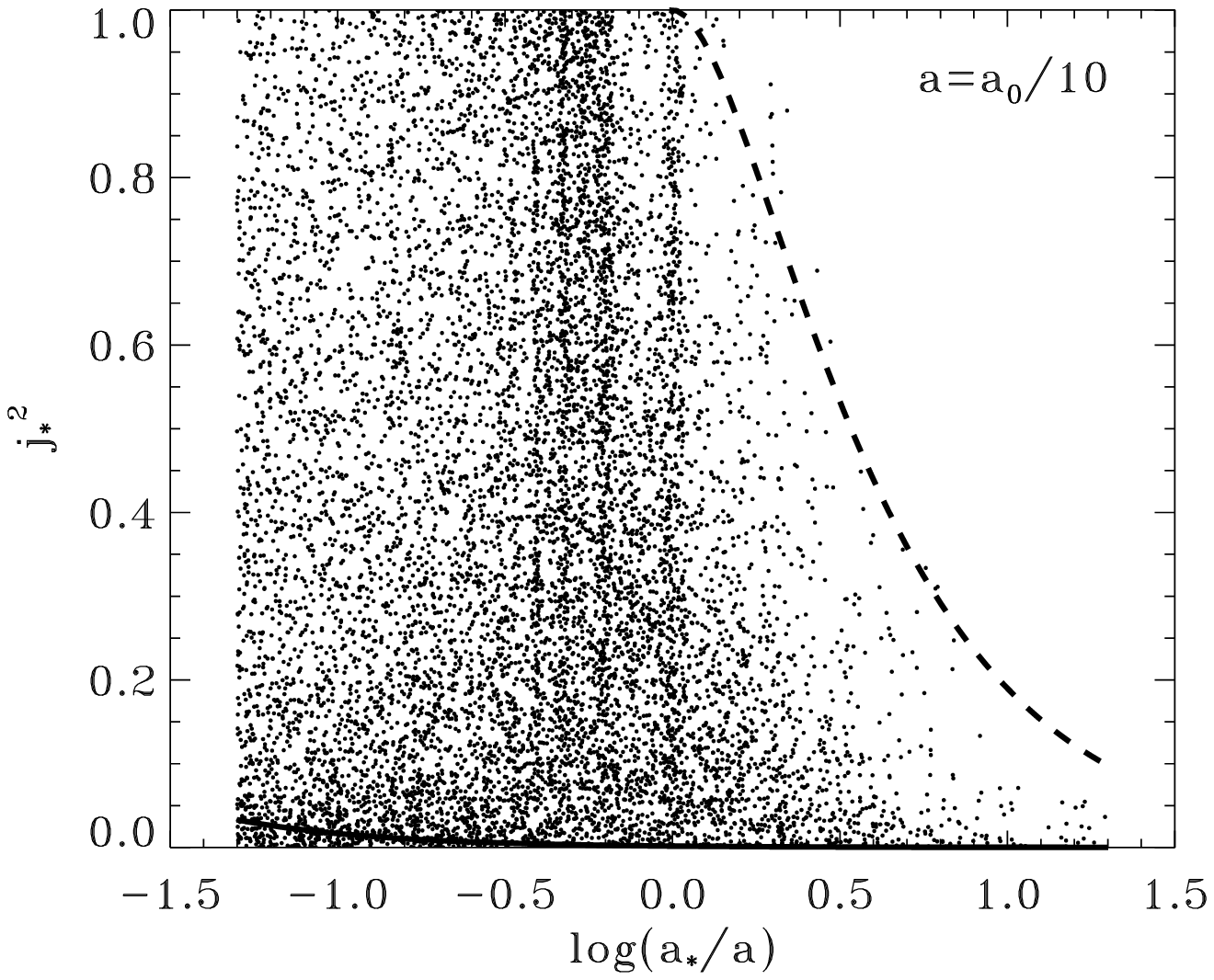}{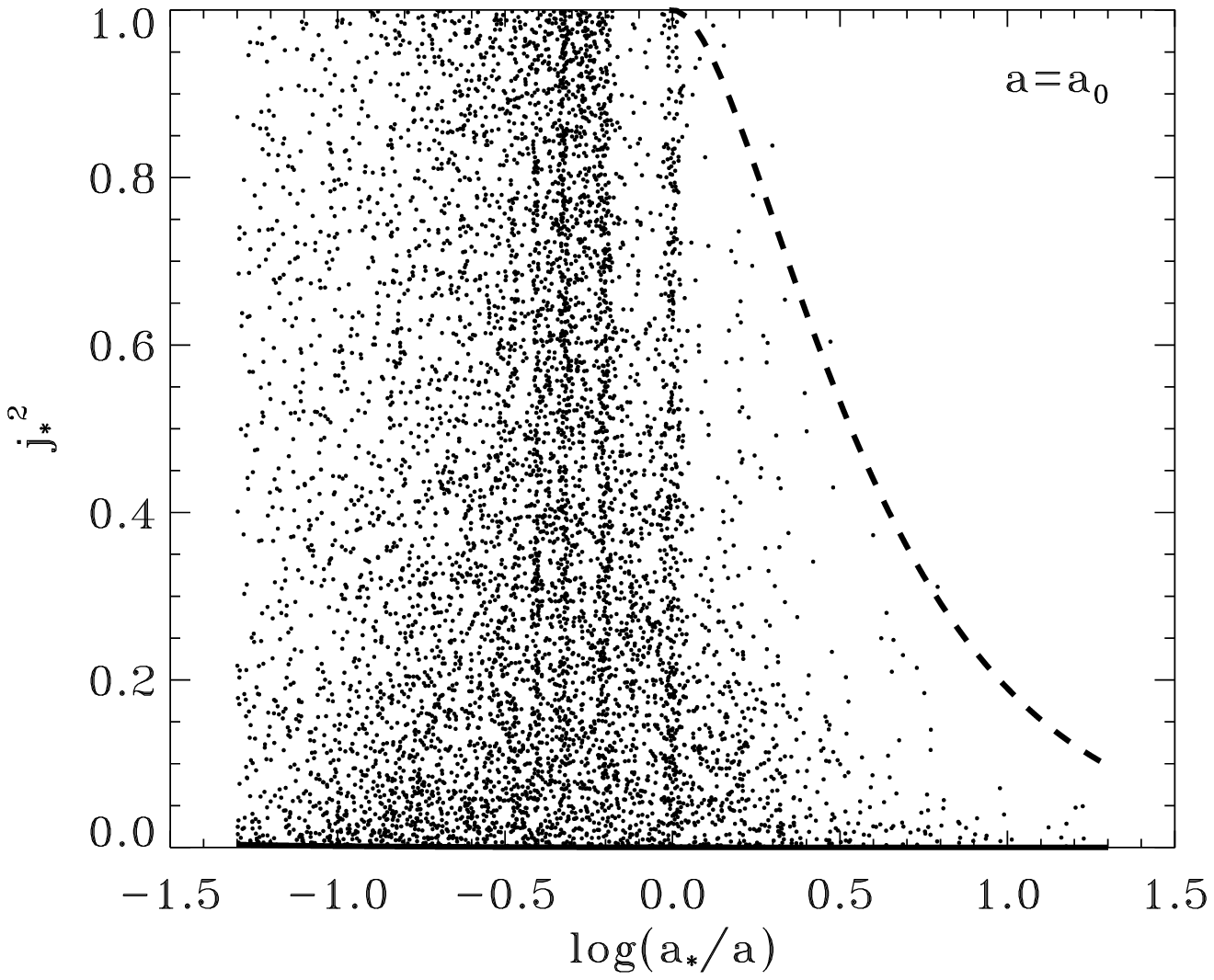}\\
\caption{Distribution of disrupted stars in the $a_*-j_*^2$ plane, assuming $e=0.1$. {\it Left panel:} $a=a_0/10$. {\it Right panel:} $a=a_0$. 
The solid and dashed lines delineate, respectively, the tidal loss cones and the interaction loss cones. The parameters of the binary-stellar
cusp are  $q=1/81$, $M_7=1$, $\sigma_{100}=1$, and $\gamma=2$. The excess of disrupted stars at radii $a_*=a(m/n)^{2/3}$, where $m,n=1,2,3...$, 
is due to resonant interactions.}
\label{disrupt}
\end{figure*}

The semimajor axis of a MBHB typically shrinks by a factor of $\sim 10$ during the process of cusp erosion via three-body scatterings \citep{sesana08}. 
Below we scale the same scattering experiments and present results for two cases, $a=a_0$ and $a=a_0/10$. Figure~\ref{disrupt} shows the distribution of 
disrupted stars in the $a_*-j_*^2$ plane for $e=0.1$. The fraction of disrupted stars exceeds 19\% in the case $a=a_0/10$, and is close
to 13\% for $a=a_0$. Many of the stars that get disrupted are initially located outside the tidal loss cone, showing that $j_*$ is not conserved 
during the three-body interaction; on the other hand, stars initially outside the interaction loss cones get disrupted only rarely. Figure~\ref{disrupt} 
also shows an excess of stars at the resonance radii $a_*=a(m/n)^{2/3}$, where $m,n=1,2,3...$, indicating the importance of resonant 
interactions in refilling the tidal loss cone.

\begin{figure*}
\plottwo{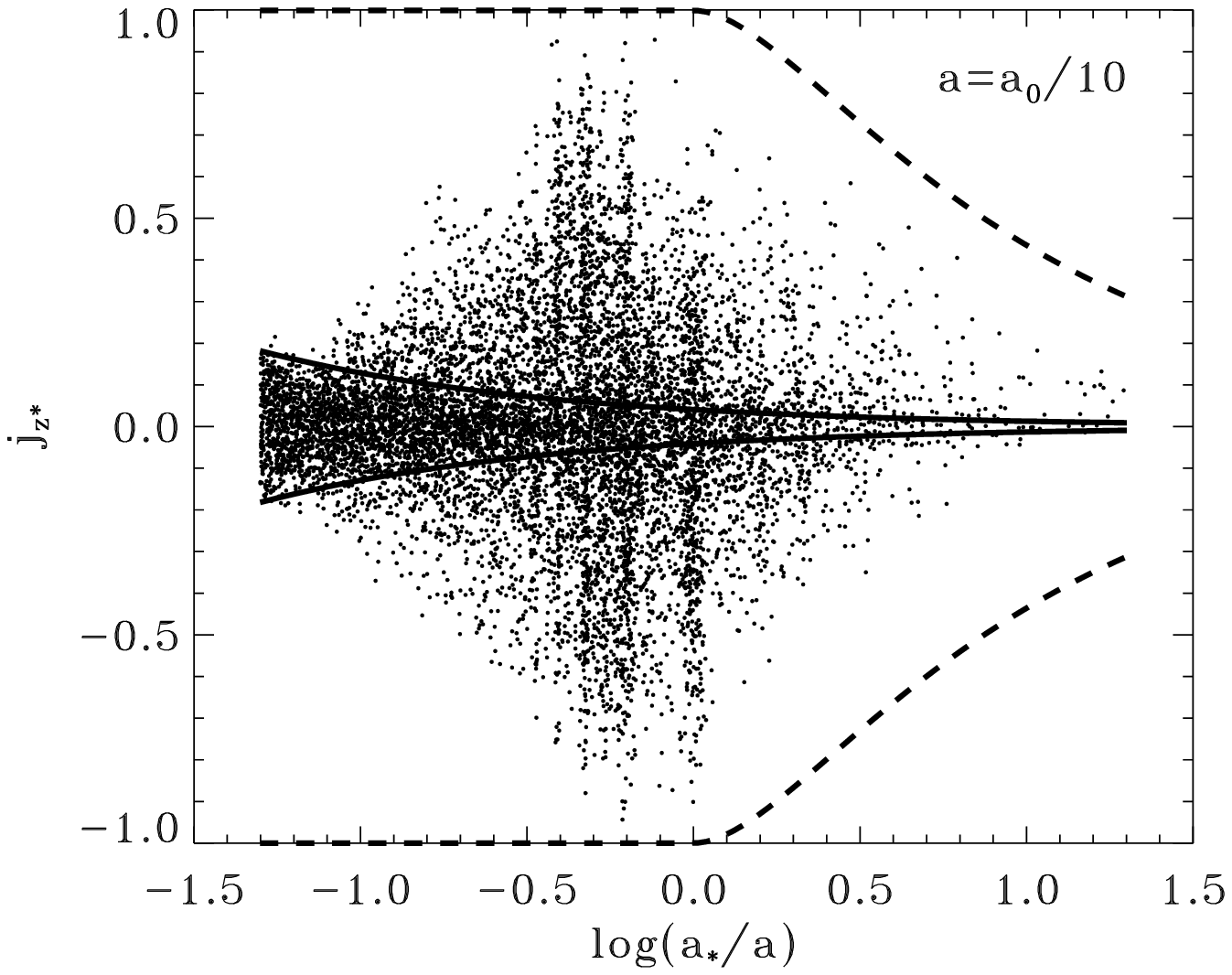}{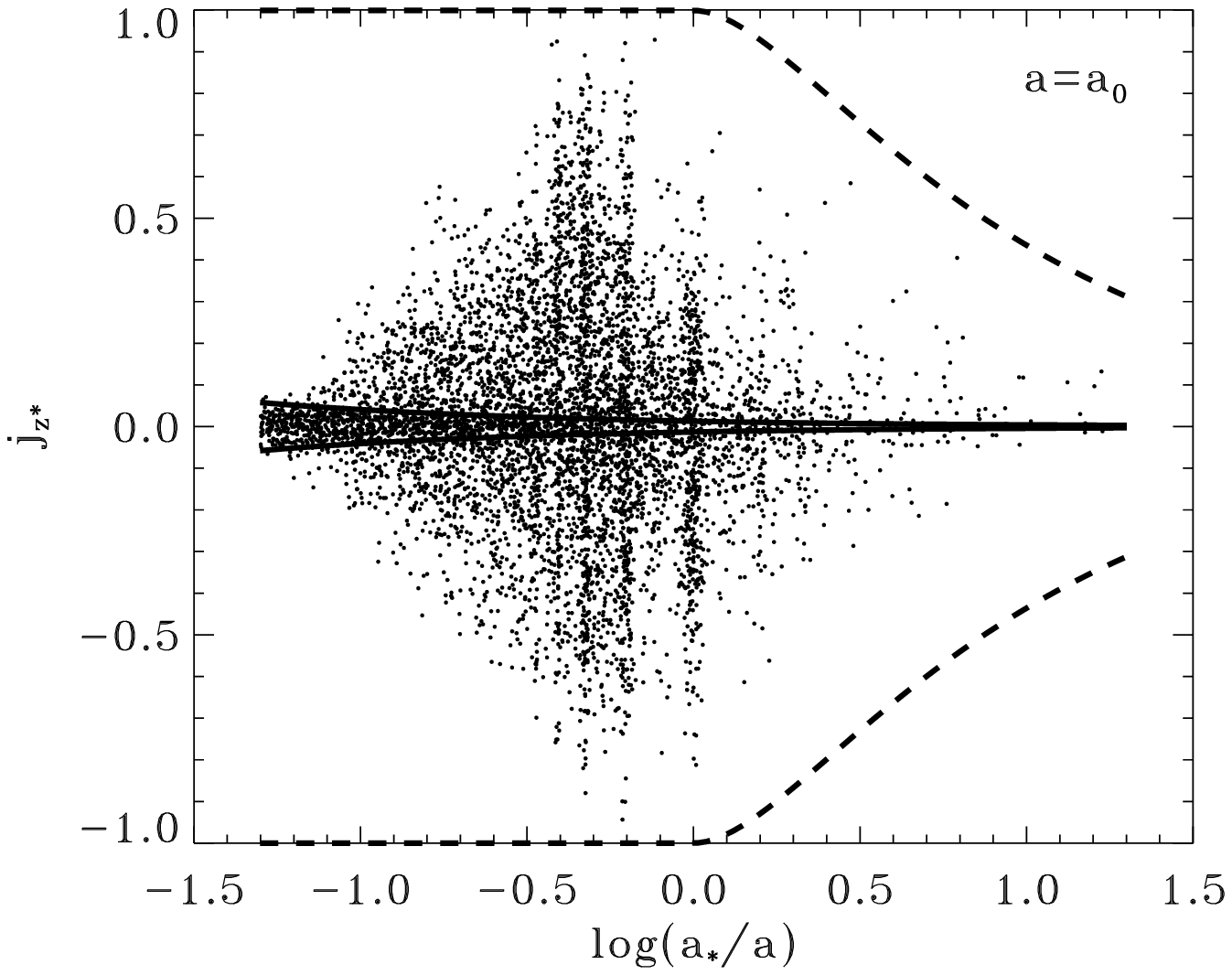}\\
\caption{Same as Fig.~\ref{disrupt} but in the $a_*-j_{z*}$ plane.}\label{jzmulti}
\end{figure*}

For a better understanding of the nature of disrupted stars we depict in Figure~\ref{jzmulti} their distribution in the $a_*-j_{z*}$ plane. 
Both theoretical and numerical studies show that for stars that lie inside the binary orbit (with semimajor axis $a_*<a/2$), the 
angular momentum component parallel to the binary orbital angular momentum, $j_{z*}$, does not change, while the angular momentum 
component perpendicular to $j_{z*}$ undergoes secular evolution \citep{kozai62,lidov62}. This implies that stars  in the wedge-like region 
$|j_{z*}|<j_{\rm lc}(r_{t1}/a_*)$ will undergo secular evolution and finally enter the tidal loss cone and get disrupted. Figure~\ref{jzmulti} confirms
that the majority of the disrupted stars with $a_*\la a/2$ have $|j_{z*}|\la j_{\rm lc}(r_{t1}/a_*)$, i.e. lie within the region delimited by the solid 
lines. When $a_*\ga a/2$, however, stars on eccentric orbits cross the orbit of the secondary hole, and can get disrupted even if 
$|j_{z*}|\gg j_{\rm lc}(r_{t1}/a_*)$. These stars are difficult to model as their orbits are chaotic. The size of the interaction loss cone 
relative to the Kozai wedge increases with $a$ ($r_{t1}/a$ decreases). As a result, strong chaotic three-body interactions rather 
than cumulative secular effects are responsible for the majority of the disruptions.  

For an isotopic stellar distribution, the fraction of stars having semimajor axis in the range $(a_*,a_*+\Delta a_*)$ that reside inside the 
``Kozai wedge'' is given by  
$$
f_K(a_*/a)=\int_{j_{\rm lc}(r_{t1}/a_*)}^{j_{\rm lc}[(1+e)a/a_*]}dj_*\int_{-j_{\rm lc}(r_{t1}/a_*)}^{j_{\rm lc}(r_{t1}/a_*)}dj_{z*} \nonumber
$$
\begin{equation}
~~~=2j_{\rm lc}(r_{t1}/a_*)j_{\rm lc}[(1+e)a/a_*]-2j_{lc}^2(r_{t1}/a_*).
\label{fk} 
\end{equation}
In a binary system with $M_7=1$, $e=0.1$, $\sigma_{100}=1$, $\gamma=2$, and $a=a_0$, the mean fractions $f_K(a_*/a)$ in the strong-interaction 
regime $a/2<a_*<2a$ are $(0.0089,0.025,0.044,0.074)$ for $q=(1/9,1/81,1/243,1/729)$. These theoretical estimates are significantly smaller than 
the tidal disruption probabilities derived in \S~\ref{effqe} except when $q=1/9$, highlighting the importance of chaotic interactions in the repopulation
of the loss cone. For $q=1/9$, the theoretical tidal disruption cross section becomes comparable to the numerical one, because stars in the 
chaotic-interaction regime are more susceptible to early ejection. 

\begin{figure*}
\plottwo{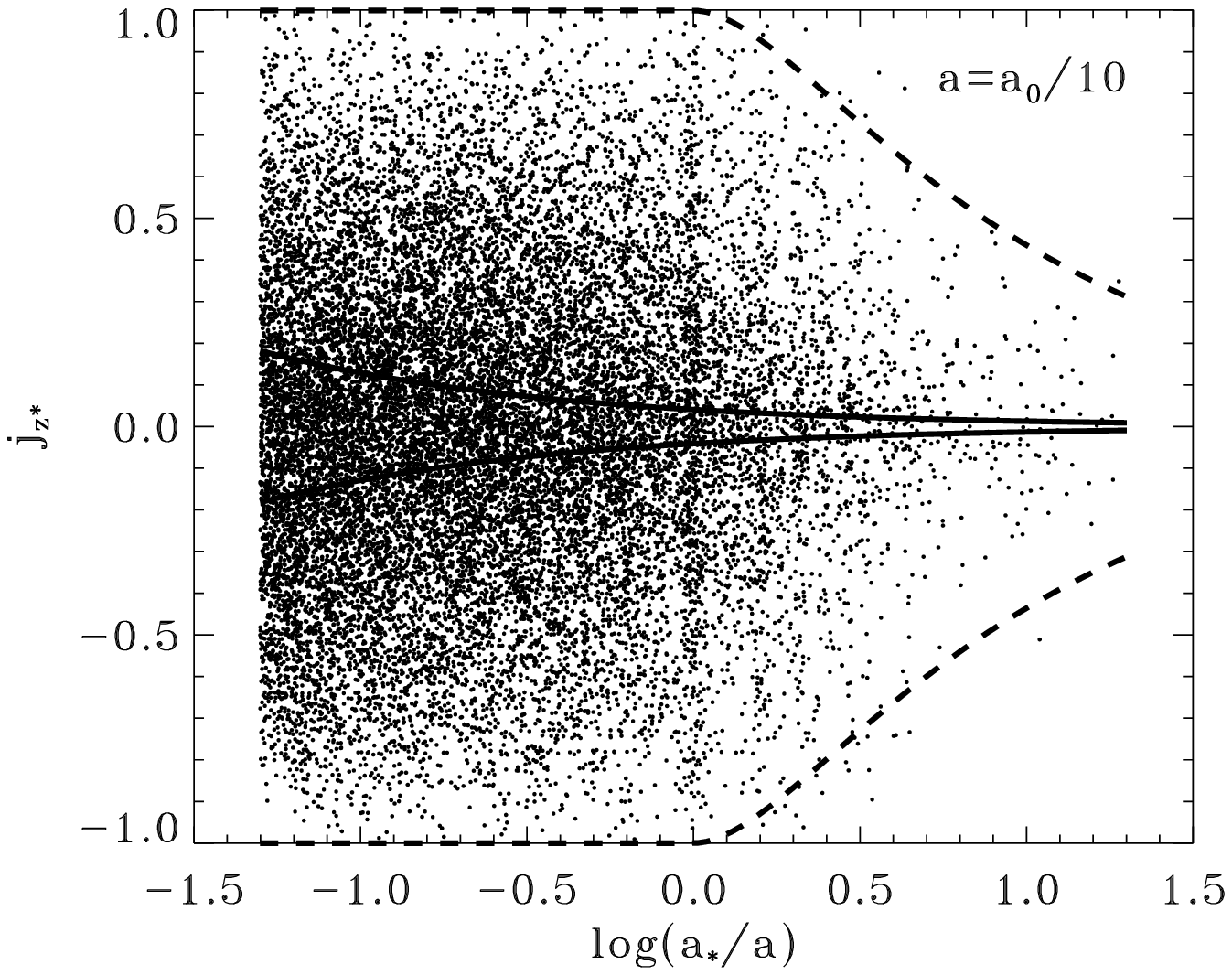}{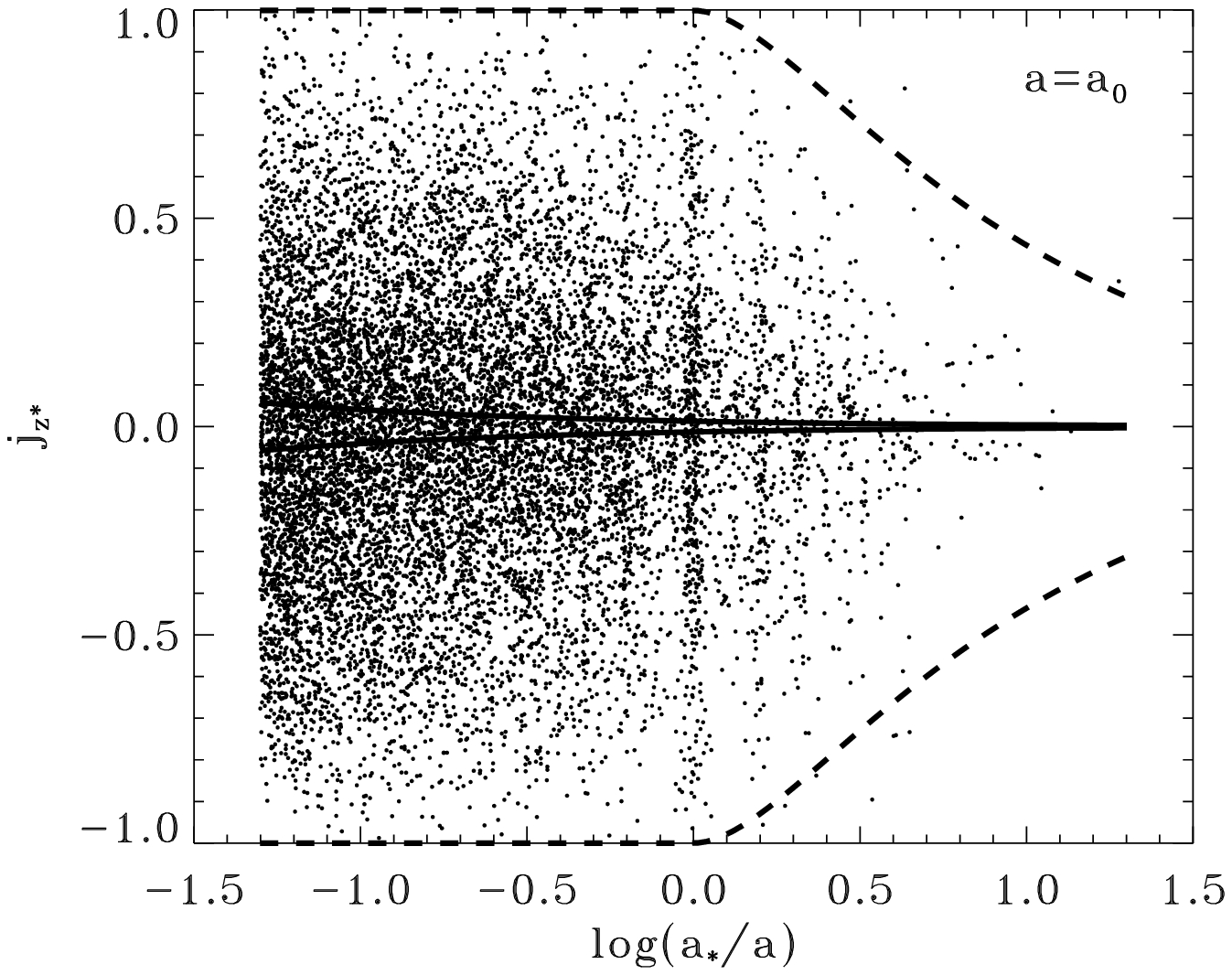}\\
\caption{Same as Fig.~\ref{jzmulti} but for $e=0.9$.}\label{phasee9}
\end{figure*}

Figure~\ref{phasee9} shows the distribution of disrupted stars in the $a_*-j_{z*}$ plane for the extreme $e=0.9$ case. Note that the interaction 
loss cone is $j_{lc}[(1+e)a/a_*]$ when $e$ is large. The number of disrupted stars increases significantly relative to the low binary eccentricity case,
to about 38\% when $a=a_0/10$ and 24\% when $a=a_0$. This enhancement occurs as more stars now cross the orbit of the secondary hole and interact 
chaotically with the binary. Moreover, any trace of the Kozai wedge disappears in this case. This is because, when $e=0.9$, the apoastron of the 
secondary hole is $0.1a$; therefore, even stars with $a_*\ll a$ experience strong interactions with the secondary hole that destroy the secular, 
coherent accumulation of the Kozai mechanism.  

\subsection{Disruption timescales}\label{tdr}

\begin{figure*}
\plottwo{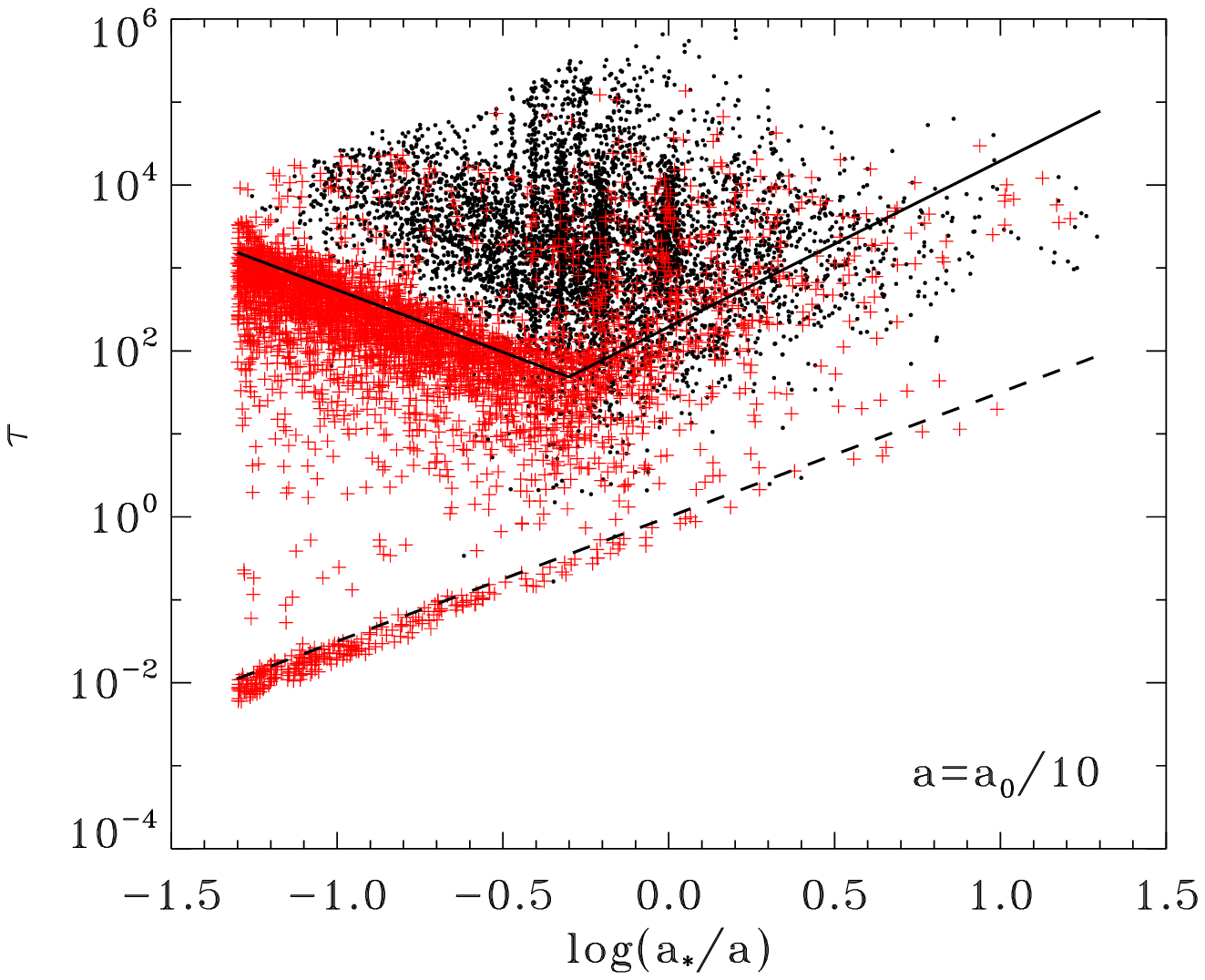}{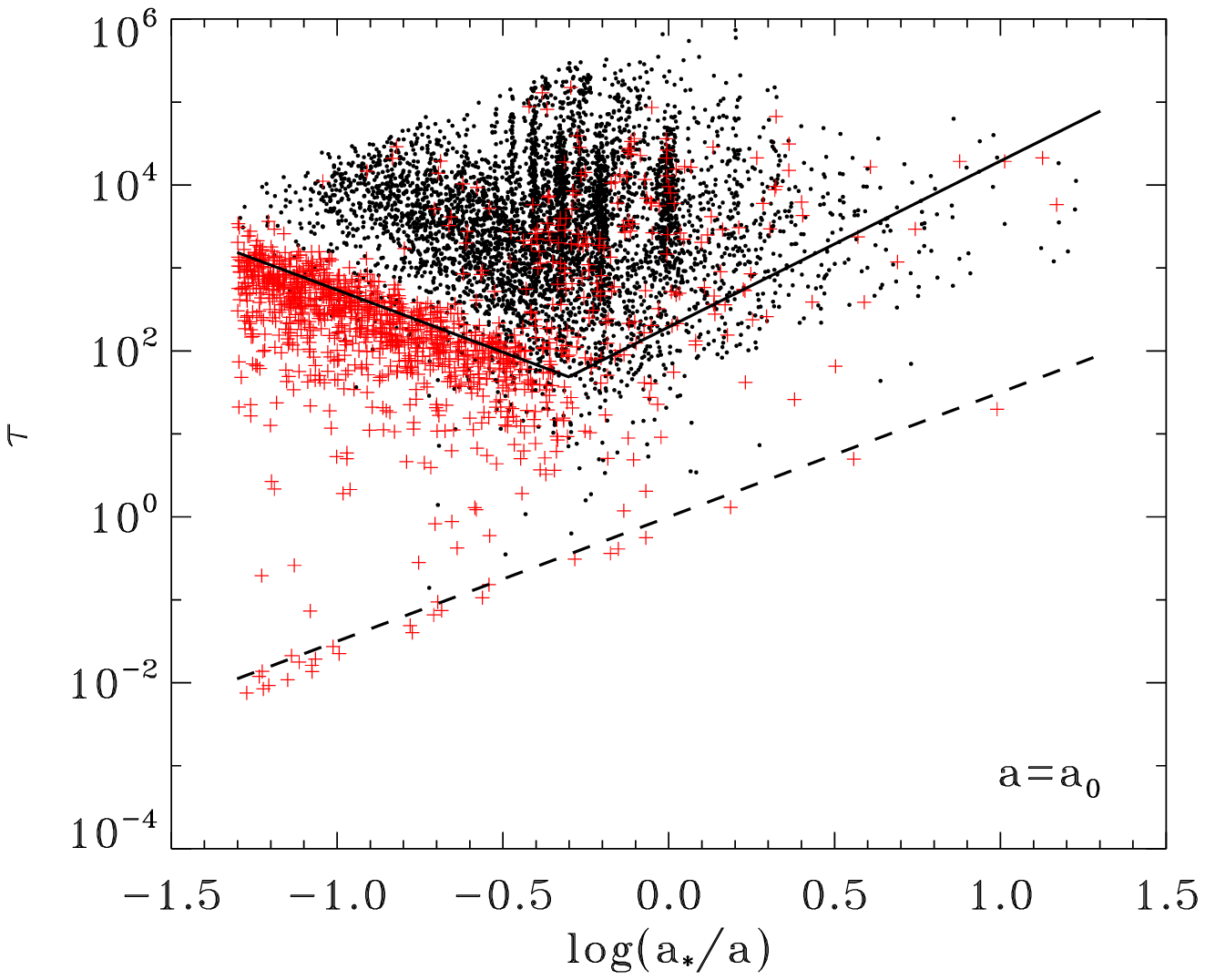}\\
\caption{Tidal disruption timescales in unit of the binary period $P$ for $a=a_0/10$ ({\it left panel}) and $a=a_0$ ({\it right panel}),
as a function of the initial semimajor axis of the intruder stars. The dashed line represent the initial stellar orbital periods, 
and the solid line marks the analytical Kozai timescale given by eq.~(\ref{tkglob}). The red crosses refer to disrupted 
stars initially inside the Kozai wedge, $|j_{z*}|<j_{\rm lc}(r_{t1}/a_*)$. System parameters as in Fig.~\ref{disrupt}.}\label{drtime}
\end{figure*}

During a scattering experiment a bound star may enter the tidal sphere of the primary black hole many times before it is ejected. When calculating the 
disruption rate, it is the time when the star first crosses $r_{t1}$ that is relevant: in the following, we refer to this time as 
the ``tidal disruption timescale''. In our numerical integrations, we record the times when each star first reaches 21 different separations 
logarithmically distributed within the range $[\log (100 r_{t1}/a_0),\log (r_{t1}/100 a_0)]$. Then, for any binary separation between $a_0/100$ and 
$100 a_0$, the tidal disruption timescale can be derived by interpolating between the recorded times. For example, the time when a star reaches $10 r_{t1}/a_0$ 
can be viewed as the tidal disruption timescale (in units of the binary orbital period) when the binary has shrunk to $a=a_0/10$. Figure~\ref{drtime} 
shows the tidal disruption timescale, $\tau$ (in unit of the binary period $P$), as a function of the initial semimajor axis, for $e=0.1$. The 
dashed line indicates the stellar orbital period, $P_*=P (a_*/a)^{3/2}$. The cross symbols clustered around the dashed line represent stars 
that are disrupted within one orbital period because their initial pericenter distances are smaller than the tidal radius of $M_1$. 
The solid lines in Figure~\ref{drtime} trace instead the Kozai timescale. The formula derived in equation (\ref{tk}) applies to stars 
that orbit close to the primary hole while the secondary is far away (``inner problem"). In this case, the quadrupole force exerted by the 
secondary on the star,
\begin{equation}
F_T\sim (Gm_*M_2a_*)a^{-3},
\end{equation}
causes a torque $a_*F_{T}$ that changes the angular momentum of the star on the timescale
\begin{equation}\label{tkprim}
T_ K'= \frac{\J_*}{d\J_*/dt}\simeq \frac{m_*(GM_{1}a_*)^{1/2}}{a_*F_{T}}=\frac{P}{2\pi q}\left(\frac{a}{a_*}\right)^{3/2}.
\end{equation}
In spite of the many simplifications in the derivation of equation~(\ref{tkprim}), $T_K'$ differs from the actual $T_K$ by only a factor of 4/3. 

Based on our understanding of the Kozai effect in the inner problem, we can now estimate the Kozai timescale for the case of orbit crossing between the
intruder star and the secondary hole (``outer problem"). When $a_*\gg a$, the quadrupole force exerted on the star by the binary is
\begin{equation}
F_{T}\sim \frac{Gm_*M_2a}{a_*^3},
\end{equation}
and $a_*F_{T}$ is the corresponding torque. Since the pericenter of a star must be smaller than $a$ (star lies in the interaction loss cone)
for it to be scattered into the tidal loss cone (see \S~\ref{phasespace}), the maximum stellar angular momentum is $m_*j_{lc}(a/a_*)\sqrt{GM_{1}a_*}$. 
The Kozai timescale in the outer problem is then given by the ratio of the maximum angular momentum and the torque,
\begin{equation}\label{tkout}
T_K\propto \frac{m_*j_{lc}(a/a_*)(GM_{1}a_*)^{1/2}}{a_*F_{T}}\propto (a_*/a)^2P.
\end{equation}
Since the transition radius between the inner and outer problems is $a_*\sim a/2$, continuity between equations (\ref{tk}) and (\ref{tkout}) 
yields
\begin{eqnarray}\label{tkglob}
T_K&=&
\left\{
\begin{array}{ll}
\frac{2}{3\pi q}\left(\frac{a_*}{a}\right)^{-3/2}P\,\,\,\,\,\,\,\,(a_*\le a/2)\\
\frac{16\sqrt{2}}{3\pi q}\left(\frac{a_*}{a}\right)^{2}P\,\,\,\,\,\,\,\,(a_*> a/2)
\end{array}.
\right.
\end{eqnarray}
In Figure~\ref{drtime} the crosses mark stars initially inside the Kozai wedge, so that crosses around the solid lines represent stars 
fed into the tidal loss cones via the Kozai mechanism. The dots mark instead stars initially outside the Kozai wedge; these participate to 
the chaotic loss-cone repopulation, and are disrupted on timescales that are typically longer than the Kozai timescale. As 
the binary orbital separation increases, less stars are disrupted by the Kozai mechanism and stars on chaotic orbits make a larger 
contribution to the repopulation of the tidal loss cones.

\begin{figure*}
\plottwo{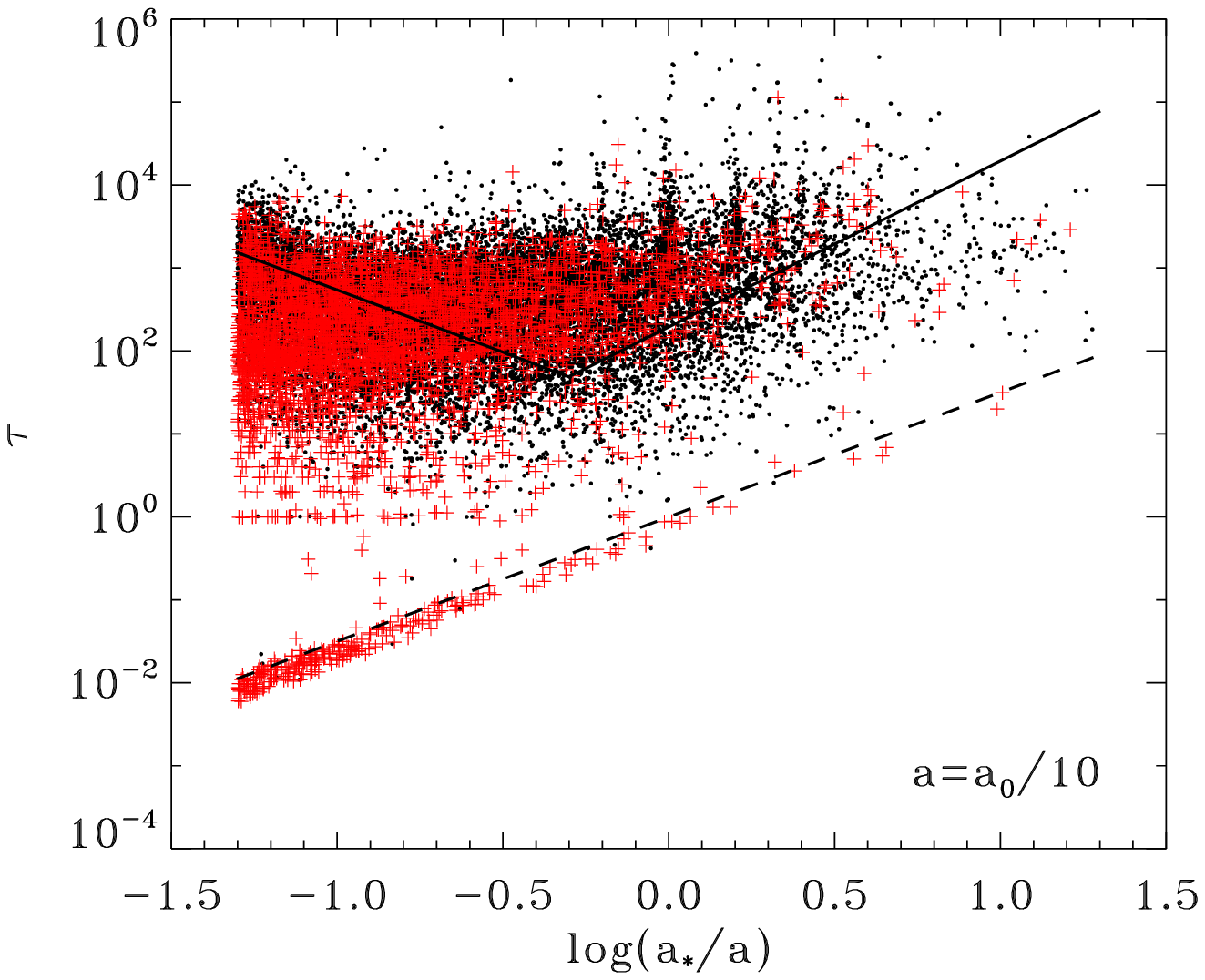}{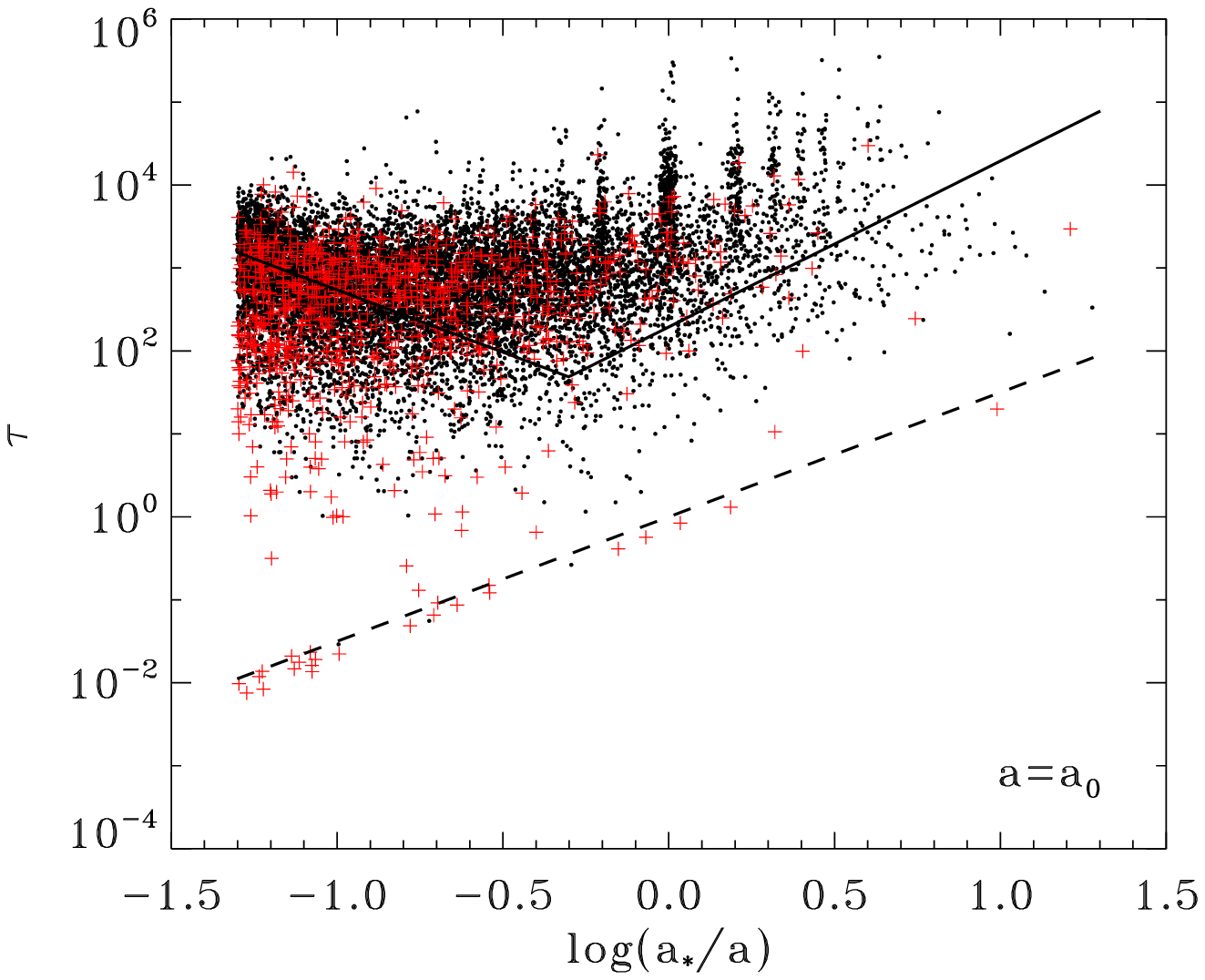}\\
\caption{Same as Fig.~\ref{drtime} but for $e=0.9$.}\label{drte9}
\end{figure*}

Figure~\ref{drte9} shows the tidal disruption timescales for the high-eccentricity, $e=0.9$ case. The cross symbols no longer cluster along
the solid lines and the dispersion is larger; this is because resonant interactions are now stronger and partially suppress the secular 
Kozai evolution.

\subsection{Relativistic and cusp-induced apsidal precession}\label{greff}

Relativistic effects close to the central MBH and the presence of an extended stellar cusp can both induce the precession of the pericenter of 
a stellar orbit. These would act to suppress the Kozai mechanism if the associated precession rates were higher than the Kozai rate.

The effect of relativistic precession can be understood as follows. The pericenter of a star with semimajor axis $a_*$ and eccentricity $e_*$ undergoing 
a secular Kozai evolution precesses at an approximate rate 
\begin{equation}\label{omegak}
\dot{\omega}_K\simeq\frac{5}{(1-e_*^2)^{1/2}}T_K(a_*)^{-1}.
\end{equation}
When $e_*$ grows to $e_*\simeq 1-r_{t1}/a_*$, the star gets tidally disrupted by the primary hole during the pericenter passage. 
Using $(1-e_*^2)a_*\simeq 2r_{t1}$ and equation~(\ref{tkglob}), equation~(\ref{omegak}) gives the Kozai precession rate at the  
tidal disruption radius,  
\begin{eqnarray}
\dot{\omega}_K&\simeq&
\left\{
\begin{array}{ll}
\frac{15\pi q}{2\sqrt{2}P(a)}\left(r_{t1}\over a\right)^{-1/2}\left(a_*\over a\right)^{2}\,\,\,\,\,\,\,\,(a_*<a/2)\\
\frac{15\pi q}{32P(a)}\left(r_{t1}\over a\right)^{-1/2}\left(a_*\over a\right)^{-3/2}\,\,\,\,\,\,\,\,(a_*\ge a/2)
\end{array}.
\right.\label{okarr}
\end{eqnarray}
The general relativistic precession rate is instead
\begin{equation}
\dot{\omega}_{\rm GR}\simeq\frac{6\pi GM_1}{(1-e_*^2)c^2a_*}P(a_*)^{-1}=\frac{3\pi}{2P(a)}\left(\frac{r_{\rm S1}}{r_{t1}}\right)\left(a_*\over a\right)^{-3/2}.
\end{equation}
The condition $\dot{\omega}_{\rm GR}=\dot{\omega}_K$ admits only one solution, $a_{*,{\rm cri}}=(8\sqrt{2}\xi)^{-2/7}a$, where
\begin{eqnarray}
\xi& \equiv &\frac{5q}{16}\left(\frac{r_{t1}}{r_{\rm S1}}\right)\left(\frac{a}{r_{t1}}\right)^{1/2}\\
&\simeq&1.64\times10^3(3-\gamma)^{1/2}M_7^{-1/3}\sigma_{100}^{-1}q^{(7-2\gamma)/(6-2\gamma)}\left(\frac{a}{a_0}\right)^{1/2}.\label{xi}
\end{eqnarray}
If $a_*<a_{*,{\rm cri}}$, the Kozai evolution is suppressed by GR precession. Note that when $\xi<1$, $\dot{\omega}_{\rm GR}>\dot{\omega}_K$, and stellar 
disruptions from the secular Kozai mechanisms are suppressed for the entire stellar population, leaving only chaotic encounters to contribute to the 
tidal disruption rate in this regime. Since the ratio between the Schwarzschild radius and the tidal radius of an MBH,
\begin{equation}
r_S/r_{t} \simeq 0.19 M_7^{2/3}(\rsun/r_*)(M_*/\msun)^{1/3},\label{rsrt}
\end{equation}
increases with hole mass, GR effects typically become important when $M_{\rm BH}>3.6\times10^6\,\msun$ (i.e. $r_t<10r_S$ for solar type stars).

\begin{figure}
\plotone{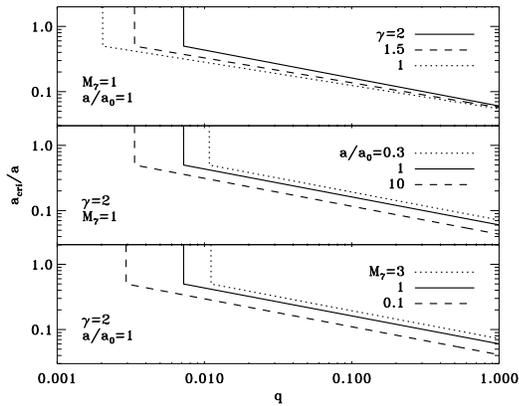}\\
\caption{Tracks of $a_{*,{\rm cri}}/a$ as a function of $q$ for different values of $\gamma$ ({\it top}), $a/a_0$ ({\it middle}), and $M_7$ ({\it bottom}). The 
variables are labeled in the top-right corner of each panel, and the fixed parameters in the lower-left corner.}\label{qacri}
\end{figure}

\begin{figure}
\plotone{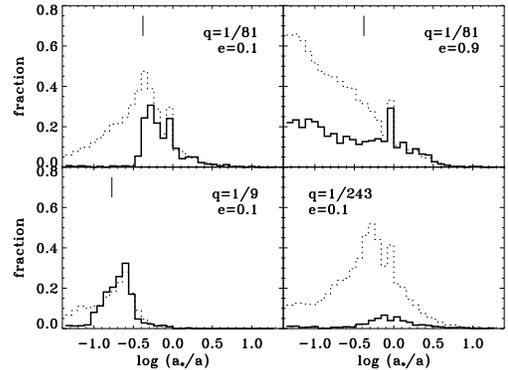}\caption{Fraction of disrupted stars in each bin of stellar semimajor axis for the GR ({\it solid histograms}) and non-GR 
({\it dotted histograms}) simulations. The short vertical lines mark the critical radius where  $\dot{\omega}_{\rm GR}=\dot{\omega}_K$. 
When $q=1/243$, $\dot{\omega}_{\rm GR}$ is always larger than $\dot{\omega}_K$.}\label{multifrac}
\end{figure}

Figure~\ref{qacri} shows the quantity $a_{*,{\rm cri}}/a$ as a function of $q$ for different combinations of the parameters $(\gamma,a/a_0,M_7)$, assuming 
$M_7\propto\sigma_*^4$ \citep{tremaine02}. Following equations~(\ref{xi}) and (\ref{acri}), the curves move toward the upper-right direction as 
$r_{t1}/r_{\rm S1}$ decreases or $r_{t1}/a$ increases. In the upper-right corner of each curve in the $q-a_{*,{\rm cri}}/a$ plane, the Kozai mechanism 
is effective, while in the lower-left corner $\dot{\omega}_{\rm GR}>\dot{\omega}_K$ and stellar disruptions are suppressed. When $q\ga 0.01$, the GR precession 
does not significantly affect the Kozai evolution of stars in the strong-interaction regime ($a_*\in[a/2,2a]$), when the stellar-disruption fraction is the 
highest (see Fig.~\ref{multifrac}). For $q\la0.01$, however, GR effects are important in the strong-interaction regime, especially in the case of
steep stellar cusps ($\gamma>2$), compact MBHBs ($a/a_0\la0.3$), or massive primary holes ($M_7\ga3$). 

To simulate numerically the effects of GR, we have run a series of scattering experiments for different values of $q$ and $e$ using the pseudo-Newtonian 
potential of \citet{paczynski80}. We integrated the equations of motion
\begin{eqnarray}
\dot{\mathbf{r}}&=&\mathbf{v}\\
\dot{\mathbf{v}}&=&-G\sum_{i=1}^{2}\frac{M_{i}(\mathbf{r}-\mathbf{r}_{i})}{|\mathbf{r}-\mathbf{r}_{i}|(|\mathbf{r}-\mathbf{r}_{i}|-r_{{\rm
  S}i})^2}\label{accgr},
\end{eqnarray}
where $r_{Si}$ is the Schwarzschild radius of the $i$th hole. Each set of experiments followed $10^4$ particles sampled in the range $a_*\in[a/20,20a]$. For 
illustrative purposes, we calculated $r_{Si}/a$ assuming $a=a_0$, $M_7=1$, $\sigma_{100}=1$, and $\gamma=2$. 
For this parameter choice, $r_{t1}/r_{S1}\simeq 5$. The integration of the stellar orbit was stopped at $1.01 r_{Si}$ to avoid the singularity at the 
Schwarzschild radius. Figure~\ref{multifrac} compares the fractions of the disrupted particles in the GR versus the non-GR simulations. When $q=1/9, 1/81$, the 
suppression of stellar disruptions by relativistic precession is important for $a_*\la a_{*,{\rm cri}}$. When $q=1/243$, $\dot{\omega}_{\rm GR}$ is always 
larger than $\dot{\omega}_K$ ($\xi=0.43$), and tidal disruptions are suppressed over the entire range of $a_*$.

\begin{figure}
\plotone{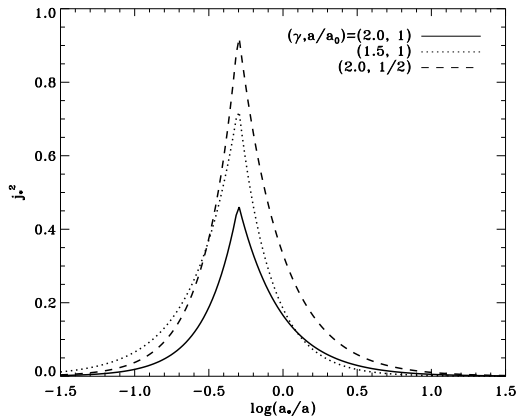}
\caption{Tracks of $j_{*,{\rm cri}}^2$ as functions of $a_*$ for different combinations of the parameters $\gamma$ and $a/a_0$. Stellar orbits 
satisfying $\dot{\omega}_c<\dot{\omega}_K$ lie below the tracks.}\label{cuspeff}
\end{figure}

The impact of an extended stellar cusp on the Kozai evolution can be addressed using similar arguments. The cusp-induced precession rate is \citep{ivanov05,merritt10}
\begin{equation}
\dot{\omega}_c=K(1-e_*^2)^{1/2}\frac{M_*(a_*)}{M_1}\frac{4\pi}{P(a)}\left(\frac{a_*}{a}\right)^{-3/2},
\end{equation}
where $K=(0.5,\sqrt{2}/\pi)$ for $\gamma=(2,1.5)$ respectively, and $M_*(a_*)$ is the stellar mass enclosed within a sphere of radius $a_*$. 
The dependence of $\dot{\omega}_c$ on $(1-e_*^2)^{1/2}$ indicates that the stellar cusp affects mostly circular orbits. Assuming the broken 
power-law distribution described in \S~\ref{background}, the condition $\dot{\omega}_c=\dot{\omega}_K$ corresponds to a critical angular momentum
\begin{eqnarray}
j_{*,{\rm cri}}^2&=&
\left\{
\begin{array}{ll}
(15/16)K^{-1}\left(\frac{a}{a_0}\right)^{\gamma-3}\left(\frac{a_*}{a}\right)^{\gamma}\,\,\,\,\,\,\,(a_*<a/2)\\
(15\sqrt{2}/356)K^{-1}\left(\frac{a}{a_0}\right)^{\gamma-3}\left(\frac{a_*}{a}\right)^{\gamma-7/2}\,\,\,\,\,\,\,(a_*\ge a/2)
\end{array},
\right.\label{j2cri}
\end{eqnarray}
above which the Kozai effect is suppressed by cusp-induced precession. Figure~\ref{cuspeff} shows tracks of $j_{*,{\rm cri}}^2$ in the $a_*-j_*^2$ plane:
once again, the phase space where the Kozai mechanism can operate is greatly reduced by cusp-induced precession. The suppression is expected to be more significant 
if $a$ or $\gamma$ increase, as the stellar mass enclosed by the stellar orbits increases.

\begin{figure*}
\plottwo{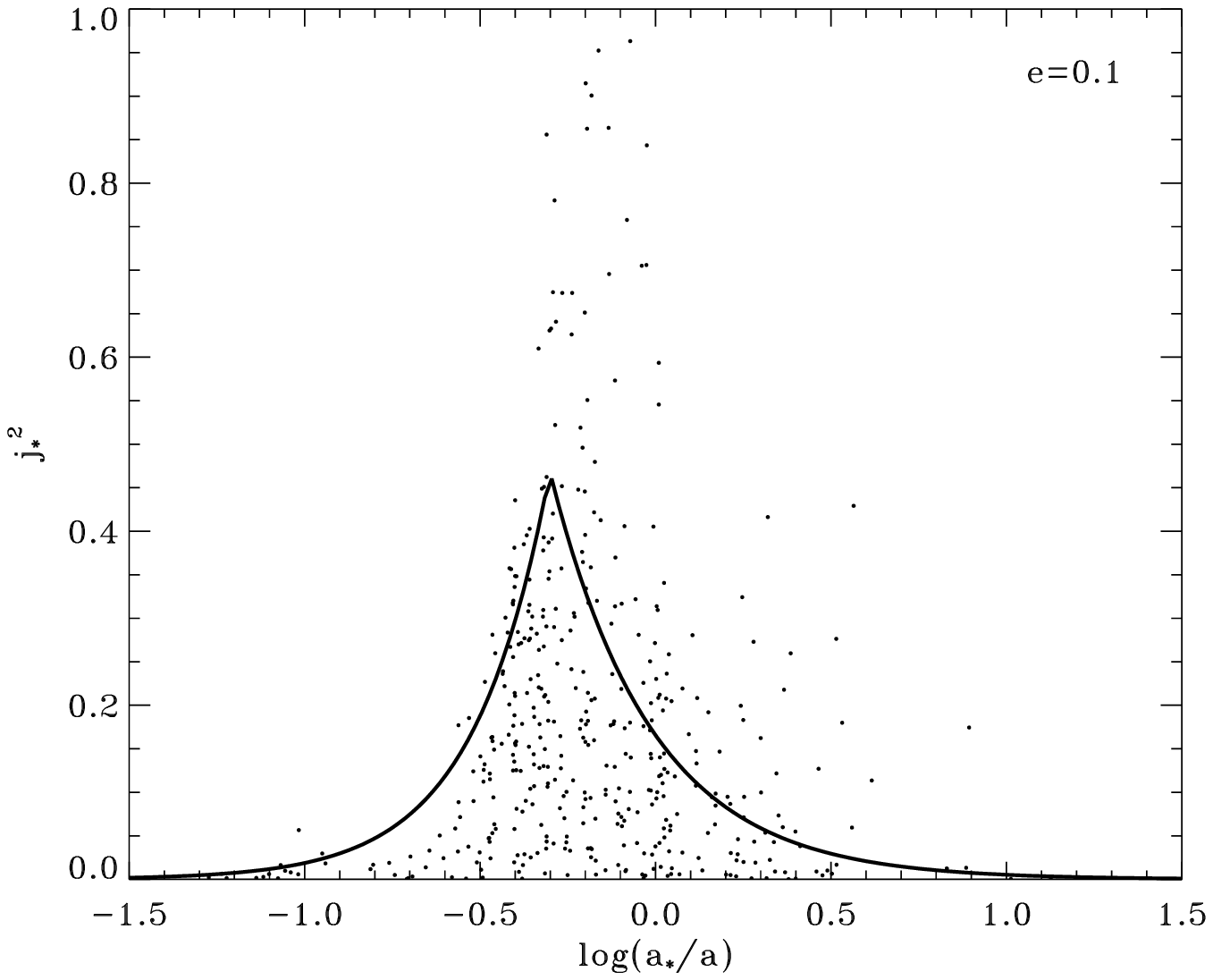}{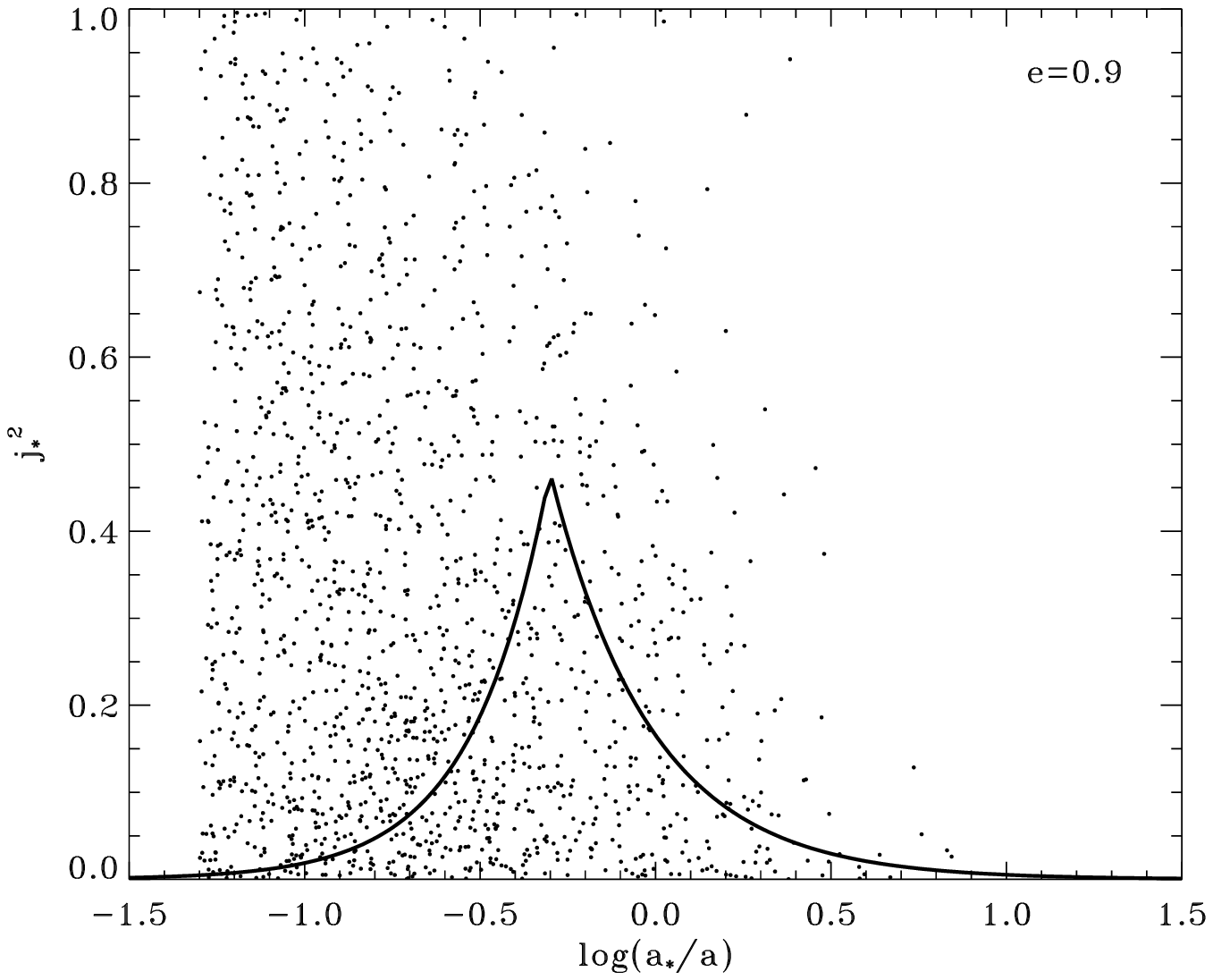}
\caption{Distribution of disrupted stars in the $a_*-j_*^2$ plane when the potential of a stellar cusp is included (see text for the parameters of 
the scattering experiments). The eccentricity of the binary is set to $e=0.1$ ({\it left}) and $0.9$ ({\it right}). The solid lines show the 
tracks of $j_{*,{\rm cri}}^2$.}\label{cuspdm}
\end{figure*}

\begin{figure*}
\plottwo{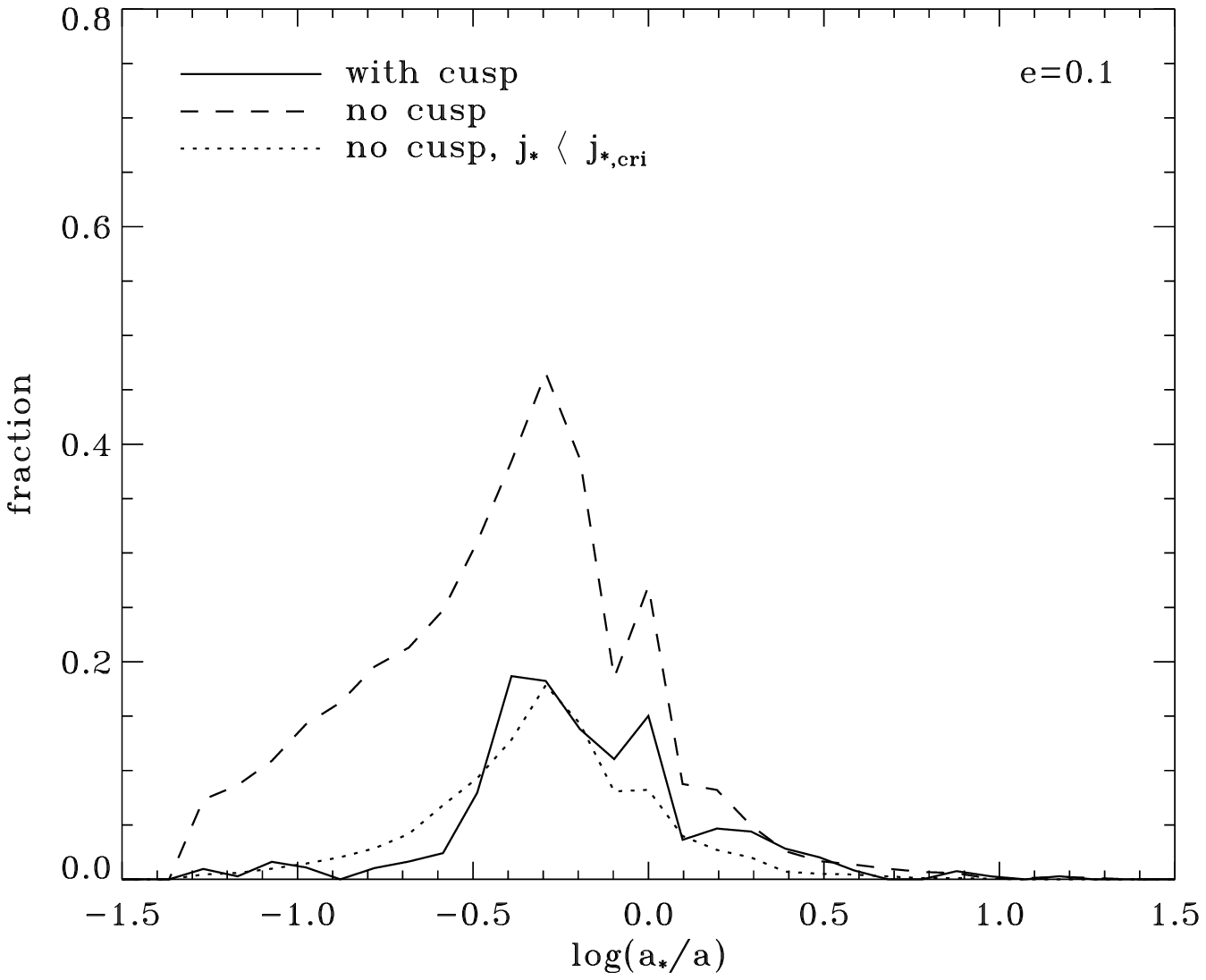}{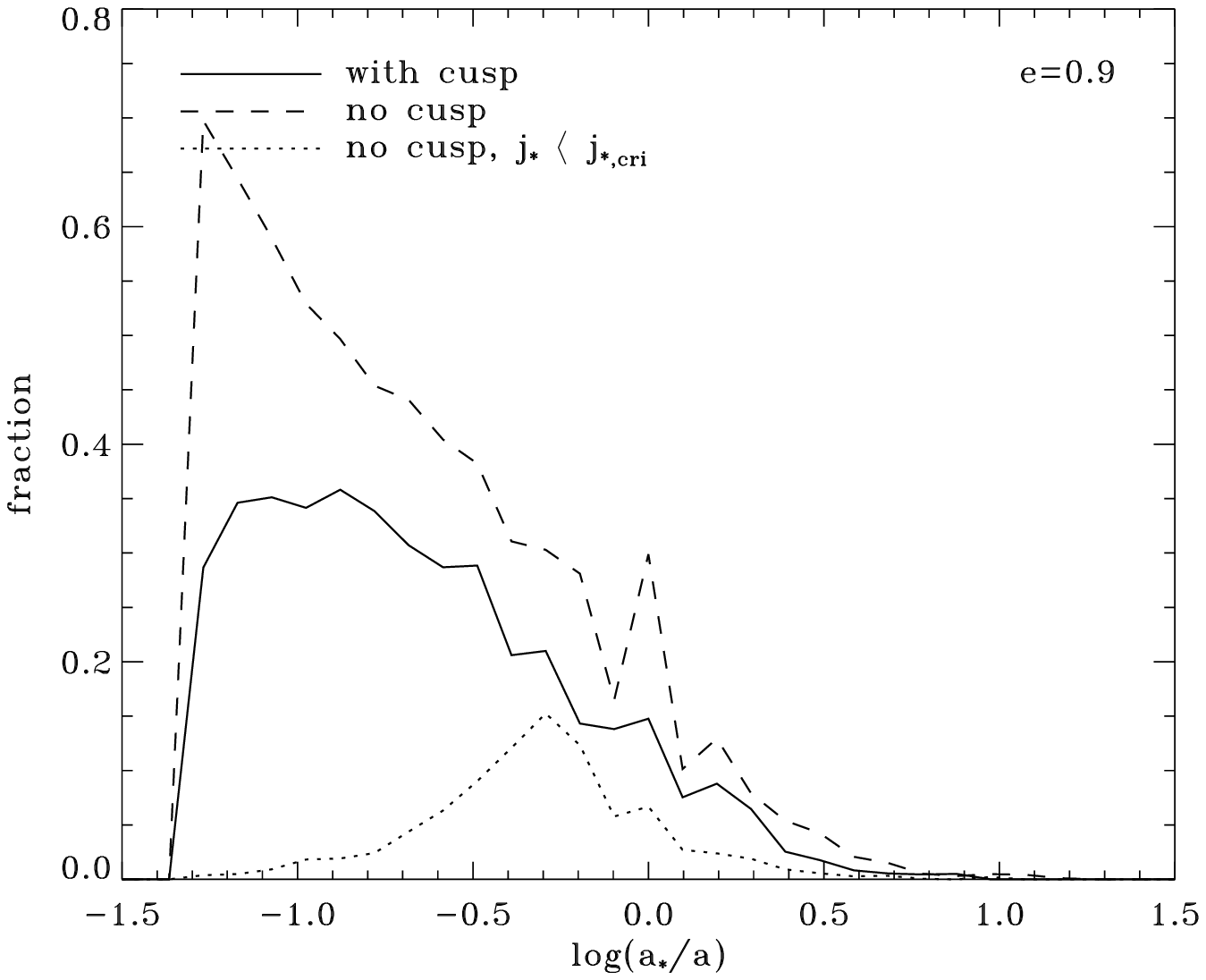}
\caption{Fraction of disrupted stars as a function of $a_*$ in scattering experiments with ({\it solid curves}) and without ({\it dashed curves}) 
the cusp potential. The dotted lines show the disrupted fractions estimated according to the $j_*<j_{*,{\rm cri}}$ criterion. The left (right) panel assumes
$e=0.1$ ($e=0.9$).}\label{cuspcom}
\end{figure*}

We mimicked the effect of a broken power-law stellar cusp (\S~\ref{background}) by including an external potential in the equations of motion. 
We set $\gamma=2$ to maximize the effect of the cusp (see Fig.~\ref{cuspeff}), and ran two sets of $10^4$ non-GR scattering experiments for $e=0.1$ and $0.9$ 
respectively. The binary parameters were set to $q=1/81$, $e=0.1$, $M_7=1$, $\sigma_{100}=1$, and $a=a_0$. Figure~\ref{cuspdm} show the initial values of
$a_*,j_*^2$ for the disrupted stars, while the fraction of tidal disruptions as a functions of $a_*$ is shown in Figure~\ref{cuspcom}. 
When $e=0.1$, tidal disruption is almost completely suppressed when $a_*\la a/2$ for stars with $j_*>j_{*,{\rm cri}}$; when $a_*\ga a/2$ a small fraction 
of stars with $j_*>j_{*,{\rm cri}}$ still get disrupted from chaotic interactions. When $e=0.9$, the suppression of stellar disruptions is also appreciable:
a large fraction of stars with $j_*>j_{*,{\rm cri}}$ are disrupted, however, because their orbits undergo chaotic intersections with the highly eccentric orbit 
of the secondary MBH. The dotted lines in Figure~\ref{cuspcom} show that, when $e$ is small, the tidal disruption fraction in the presence of a cusp 
potential is approximately equal to the fraction of disrupted stars with $j_*<j_{*,{\rm cri}}$ when the cusp is neglected, with an error that is less 
than a factor of two. When $e$ is large, however, this method severely underestimates the true disruption fraction because chaotic three-body 
interactions at $j_*>j_{*,{\rm cri}}$ are dominant.

\section{DISRUPTION RATES FOR STATIONARY BINARIES}\label{lcfr}

Having set the properties of the disrupted stars and tested the limitations of our approximations, we can now proceed to the calculation of the disruption rates. 
We fix the orbital separation $a$ and the eccentricity $e$ of the MBHB, and assume that the stellar cusp surrounding $M_1$ is isotropic and composed of 
solar-type stars only. The initial stellar distribution function, $f_0(a_*,j_*,j_{z*})\equiv dn_0/da_*$, can then be written as
\begin{equation}\label{n0}
{dn_0\over da_*}=\frac{2(3-\gamma)}{a_0}{M_2\over \msun}\left(\frac{a_*}{a_0}\right)^{2-\gamma},
\end{equation}
where $dn_0/da_*$ is the number of stars per unit semimajor axis $a_*$, and the right hand side follows from the definition of 
$a_0$. We do not consider any sharp cutoff in the stellar distribution caused by stellar collision at small radii. Collisions will likely result 
in a shallower inner density profile rather then a well defined cutoff \citep{fb02}. Moreover, the inner cusp may be repopulated by efficient gas 
inflow-induced star formation during galaxy mergers \citep{zier06}. Here we assume $\gamma=2$ and $1.5$ to account for this uncertainty. 

The stellar disruption rate at time $t$ can then be calculated from
\begin{equation}\label{ndota}
\dot{N}_*(t)=\int F(a_*/a,t)(dn_0/da_*)da_*,
\end{equation}
where $F(a_*/a,t)dt$ is the fraction of stars with semimajor axis $a_*$ that are disrupted in the time interval $(t,t+dt)$. If 
the loss-cone refilling is entirely due to the Kozai effect, then $F(a_*/a,t)$ can be derived analytically as \citep{ivanov05}
\begin{equation}\label{fda}
F(a_*/a,t)={f_K(a_*/a)\over T_K(a_*/a)}\exp(-t/T_K),
\end{equation}
where $f_K$ and $T_K$ are given by equations (\ref{fk}) and (\ref{tkprim}). The chaotic nature of strong three-body scattering events
prevent the possibility of a simple analytical description; therefore, the total disruption rate, including the contribution of 
resonant interactions, has to be computed numerically from scattering experiments. We divide the $(a_*/20a,20a_*/a)$ interval into 52 
equal logarithmic bins. Given the parameters $e$ and $r_{t1}/a$, we derive the function $F_i(\tau)$ numerically for each $a_*/a$ bin 
using the recorded tidal disruption timescales in the corresponding scattering experiments, so that $F_i(\tau)\Delta \tau$ is the fraction 
of stars in the $i$th bin that are disrupted in the time interval $(\tau,\tau+\Delta\tau)$. When deriving $F_i(\tau)$, stars sampled 
in the range $j_*(a_i)<j_{\rm lc}(r_{t1}/a_i)$ are excluded. The total stellar disruption rate at time $t$ is then given by
\begin{equation}\label{ndotn}
\dot{N}_*(t)=\sum_{i=1}^{52} \frac{F_i(t/P)(dn_0/da_*)(a_i)\Delta a_i}{P},
\end{equation}
where $a_i$ and $\Delta a_i$ are the central semimajor axis and the width of the $i$th bin. To calculate the numerical 
stellar disruption rate, one must specify $P$ and $dn_0/da_*$ in physical units. From the definition of $a_0$ and equations~(\ref{period})
and (\ref{n0}), we have
\begin{equation}
P \simeq (3\times 10^5~{\rm yr})~{M_7\over \sigma_{100}^3}~Q~(3-\gamma)^{3/2}\left(\frac{a}{a_0}\right)^{3/2}
\end{equation}
with $Q\equiv q^{3/(6-2\gamma)}/(1+q)^{1/2}$, and 
\begin{equation}
{dn_0\over da_*}\simeq (4.3\times10^6~{\rm pc^{-1}})~q^\alpha\sigma_{100}^2\left(\frac{a_*}{a_0}\right)^{2-\gamma}
\label{n0n}
\end{equation}
with $\alpha \equiv (2-\gamma)/(3-\gamma)$.
\begin{figure}
\plotone{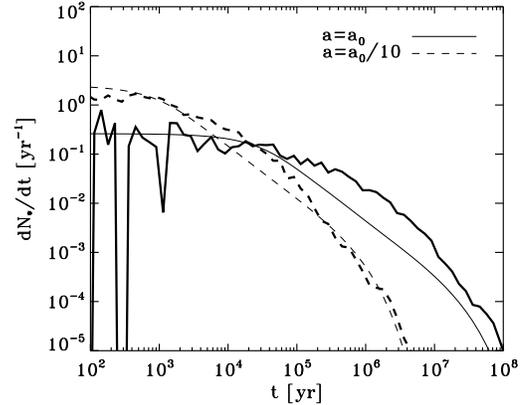}\\
\caption{Stellar disruption rates as a function of time for a stationary MBHB with semimajor axis $a=a_0/10$ 
({\it solid curves}) and $a=a_0$ ({\it dashed curves}). The thick lines are the numerical rates derived from scattering experiments, and 
the thin lines are the analytical rates. System parameters are the same as in Fig.~\ref{disrupt}, and yield $P\simeq (400,13)$ yr for 
$a=(a_0,a_0/10)$. Fluctuations in the numerical rates at early times are due to poor statistics. 
}
\label{stat}
\end{figure}
Figure~\ref{stat} shows the total stellar disruption rates calculated from equation~(\ref{ndotn}) for a MBHB with $a=a_0$ and $a=a_0/10$ (and the standard
system parameters $q=1/81$, $e=0.1$, $M_7=1$, $\sigma_{100}=1$, and $\gamma=2$). The stellar disruption rate remains constant for a timescale $P/q$ 
before decreasing rapidly, consistent with the Kozai time scaling. Compared to the rates for single MBHs fed by two-body relaxation, typically 
$10^{-4}-10^{-5}\,\ndot$, the rates on the plateau are orders of magnitude higher. As $a$ decreases, the plateau rate increases while its 
duration shortens. Since the Kozai timescale increases as $a^{3/2}$ and the number of stars enclosed in the Kozai wedge scales as $a^{1/2}$,
if $a$ shrinks by a factor of $10$ the plateau phase becomes a factor $10^{3/2}$ shorter, while the disruption rate at the plateau increases by a 
factor of $10$. The figure also depicts the analytical disruption rates calculated with equation~(\ref{ndota}) for comparison. These agree very 
well with the numerical results during the plateau phase, indicating that the tidal loss-cone refilling is initially dominated by the Kozai mechanism.
At later times, stars inside the Kozai wedge are mostly depleted, and the analytical and numerical rates start to deviate from each other. 
Deviations in the post-plateau phase increase with increasing binary orbital separations. According to our numerical calculations, about 
$(1.8\times10^4,1.1\times10^5)$ stars with $a/20<a_*<20a$ are disrupted over $10^8$ years by binaries with $a=(a_0/10,a_0)$. The
corresponding numbers in the analytical approximation are $(1.1\times10^4,3.5\times10^4)$. The difference highlights the importance of 
close, resonant encounters with the secondary hole, which change the stellar orbits in a chaotic manner and fuel the tidal loss cone.
Figure~\ref{multistat} shows the dependence of the disruption rate on the parameters $e$ and $\gamma$. Increasing the binary eccentricity 
only affects the rate in the post-plateau, chaotic-interaction-dominated phase. 
\begin{figure}
\plotone{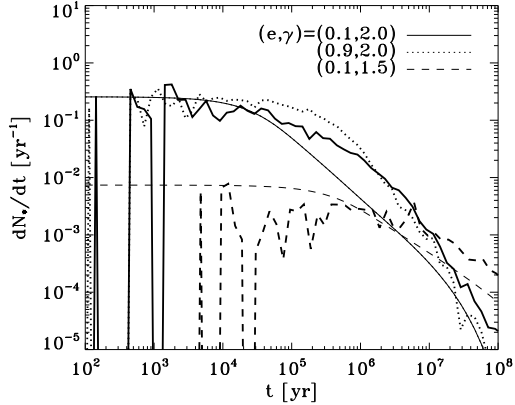}\\
\caption{Numerical ({\it thick curves}) and analytical ({\it thin curves}) stellar disruption rates as a function of time for different 
combinations of the parameters $e$ and $\gamma$ (see text for details). The semimajor axis of the binary is $a=a_0$, and all other system 
parameters are the same as in Fig.~\ref{disrupt}.}\label{multistat}
\end{figure}

The above results show that, whereas chaotic scatterings dominate the total number of disrupted stars, the Kozai theory provides a reasonably 
good description of the disruption rate during the plateau phase, as well as the correct order of magnitude of the total number of disruption. 
We can then use the Kozai theory to predict the scaling of the plateau rate with the system parameters,
\begin{equation}\label{ndotscale0}
\dot{N}_*\propto {M_*f_K\over T_K}.
\end{equation}
Here $M_*$ is total mass of the interacting stars, $M_*\propto M_1 q (a/a_0)^{3-\gamma}$ (from the definition of $a_0$), $f_K$ is the fraction of 
stars in the Kozai wedge, $f_K\propto (r_{t1}/a)^{1/2}$, and $T_K$ is the Kozai timescale, $T_K\propto q^{-1}a^{3/2}M_1^{-1/2}$ (assuming that $a_*\sim a$). Substituting into equation (\ref{ndotscale0}), and using the definitions of $r_{t1}$ and $a_0$, 
we finally obtain in the limit $q\ll 1$:  
\begin{eqnarray}\label{ndotscale}
\dot{N}_*&\propto&(3-\gamma)^{-2}q^{(4-2\gamma)/(3-\gamma)}\left({a\over a_0}\right)^{1-\gamma} M_1^{-1/3}\sigma_*^4\nonumber\\
&\propto& (3-\gamma)^{-2}q^{(4-2\gamma)/(3-\gamma)}\left({a\over a_0}\right)^{1-\gamma} M_1^{2/3},
\end{eqnarray}
where we have adopted $M_1\propto\sigma_*^4$ \citep{tremaine02} in the second proportionality. According to equation~(\ref{ndotscale}), when $q=1/81$, 
as $\gamma$ decreases from $2$ to $1.5$, the peak stellar disruption rate drops by a factor of $40$, consistent with the rates in 
Figure~\ref{multistat}. If $\gamma=2$, the disruption rate is proportional to $a^{-1}$, consistent with the left panel of Figure~\ref{stat}. 
The above analysis also implies that, when the stellar density profile is as steep as $\gamma\simeq2$, the peak stellar disruption rate is not 
sensitive to $q$, as shown by \citet{chen09}. Assuming the $M_1-\sigma_*$ relation, the peak rate should be proportional to $M_1^{2/3}$.

\begin{figure}
\plotone{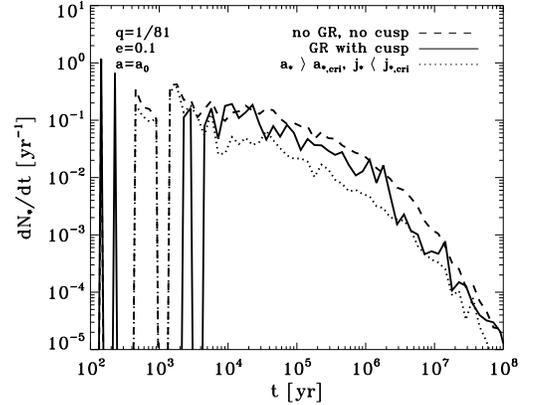}
\caption{Stellar disruption rates as a function of time for a stationary MBHB with semimajor axis $a=a_0$. Binary parameters are the same as 
those in Fig.~\ref{stat}. The solid line is derived from the scattering experiments where both GR and cusp effects are included, while the dashed 
line from the experiments when both are neglected. The dotted line is also derived from no-GR/no-cusp experiments, but only using 
stars with $a_*>a_{*,\rm cri}$ and $j_*<j_{*,{\rm cri}}$.}\label{drcom}
\end{figure}

To investigate numerically the impact of GR and cusp-induced precession on the stellar disruption rate, we ran an additional set of $10^4$ 
scattering experiments that included the two effects simultaneously (see \S~\ref{greff} for details). The binary parameters were set to $q=1/81$, $e=0.1$, 
$\gamma=2$, $M_7=1$, $\sigma_{100}=1$, and $a=a_0$. The resulting stellar-disruption rate is shown in Figure~\ref{drcom} with the solid line, and is
compared to the case in which the two effects are neglected (the dashed line). Because chaotic scatterings are not suppressed by secular effects, the two curves
differ by only about a factor of two. The contribution from chaotic scatterings in the GR$+$cusp experiments can be gauged by the difference between the 
solid and the dotted curves, the latter derived by considering only stars with $a_*>a_{*,\rm cri}$ and $j_*<j_{*,{\rm cri}}$ in the no-GR/no-cusp experiments. 
For binaries with larger $q$, larger $e$, smaller $\gamma$, or larger $a/a_0$,  stars in the chaotic-interaction regime contribute more to the disrupted 
stellar fraction, and the GR/cusp-induced suppression of tidal disruptions is milder. Disruptions rates derived by scattering experiments that neglect GR and cusp
effects, together with the analytic scalings given by equation (\ref{ndotscale}), can therefore be considered valid for binaries with $q>0.01$. 

\section{DISRUPTION RATES FOR DECAYING BINARIES}\label{sdr}

Recent calculations based on scattering experiments and ignoring stellar disruptions have suggested that both the binary semimajor axis and 
eccentricity will evolve rapidly during three-body interactions with ambient bound stars  \citep{sesana08}. On the one hand, in a shrinking MBHB 
the interaction loss cone, the tidal disruption timescale, and the cusp stellar distribution all change with time, and this evolution will affect
the stellar disruption rate. On the other hand, tidal disruptions halt the exchange of energy and angular momentum between the stars and the binary, 
altering its dynamical evolution. In order to compute a more realistic tidal consumption rate, a hybrid model is required that takes into account 
stellar ejections as well as disruptions, and that solves for the time evolution of the binary-stellar cusp system.

In evolving MBHB systems, the ratios $r_{\rm S1}/a$ and $M_*(a)/M_1$ vary with time. In this case, simulating GR and cusp effects becomes 
extremely time consuming, because additional scattering experiments need to be carried out whenever $r_{\rm S1}/a$ or $M_*(a)/M_1$ changes. For 
this reason, in this section we do not consider GR and cusp effects, and use the Newtonian scattering experiments without cusp potential to calculate 
the stellar disruption rate. We restrict the following discussion to the case $q\ga1/0.01$, where GR and stellar cusp precession suppress 
the stellar disruption rate by only a factor of two. When $q\ll0.01$, our test calculations with $q=1/729$ show that GR and cusp effects 
can suppress the stellar disruption rate by as much as a factor of ten.

\subsection{Fate of interacting stars}

\begin{figure*}
\plottwo{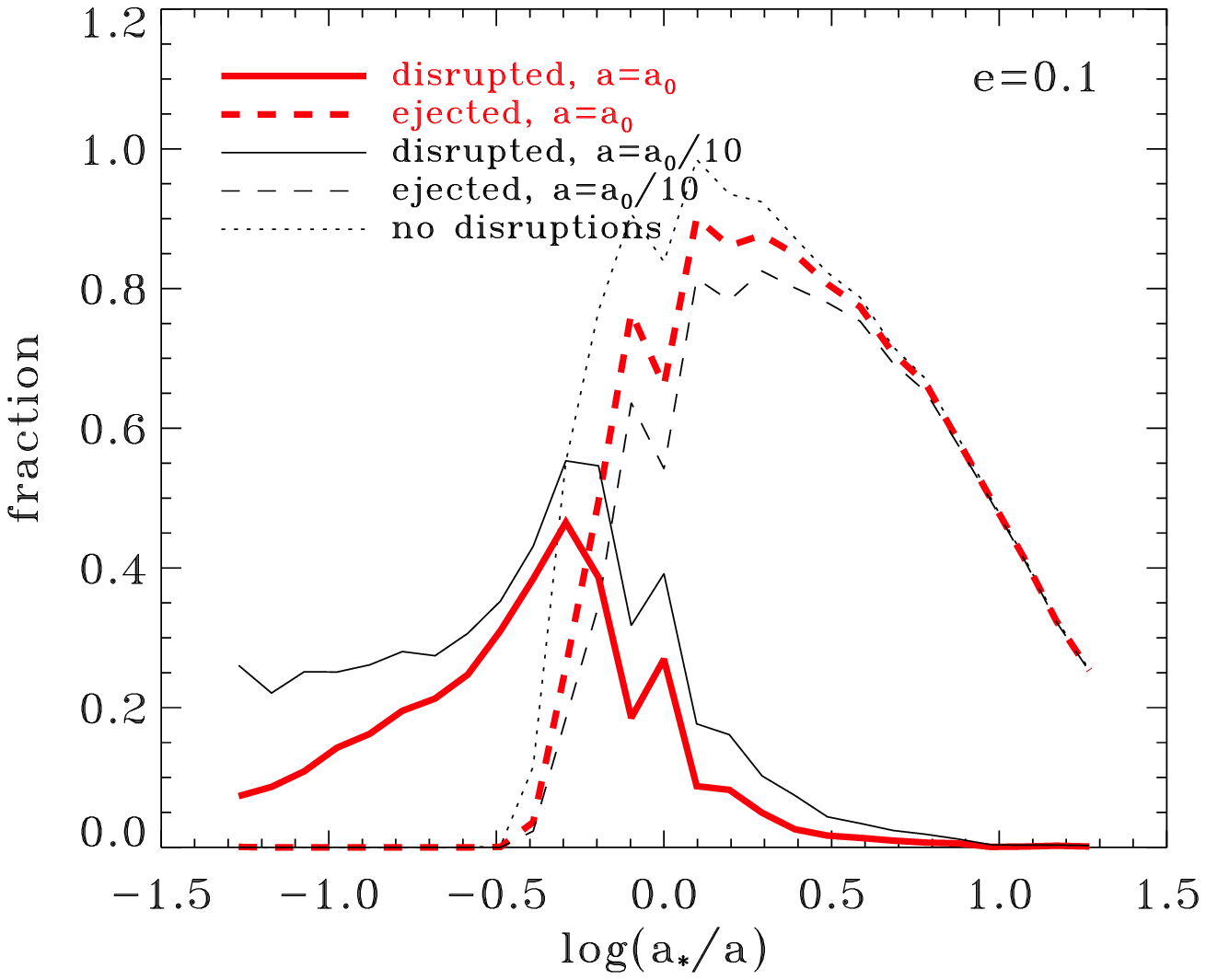}{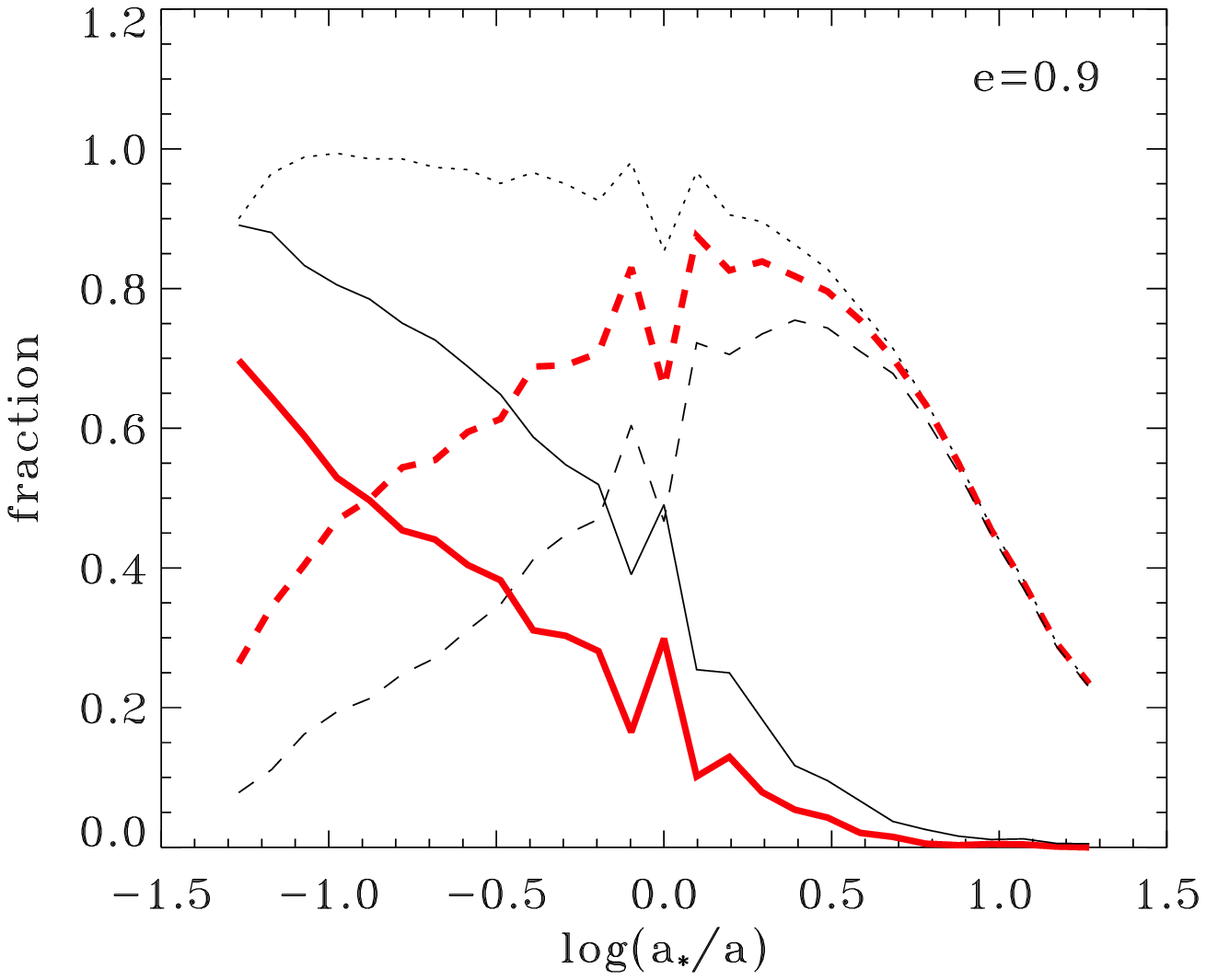}\\
\caption{Fractions of disrupted ({\it solid lines}) and ejected ({\it dashed lines}) stars as a function of stellar semimajor axis $a_*$, 
for $e=0.1$ ({\it left panel}) and $0.9$ ({\it right panel}). {\it Black thin lines:} $a=a_0/10$. {\it Red thick lines:} $a=a_0$. 
The dotted line shows the ejection fractions if disruptions are not taken into account. All other system parameters are as in 
Fig.~\ref{disrupt}.}
\label{frac}
\end{figure*}

\begin{figure*}
\plottwo{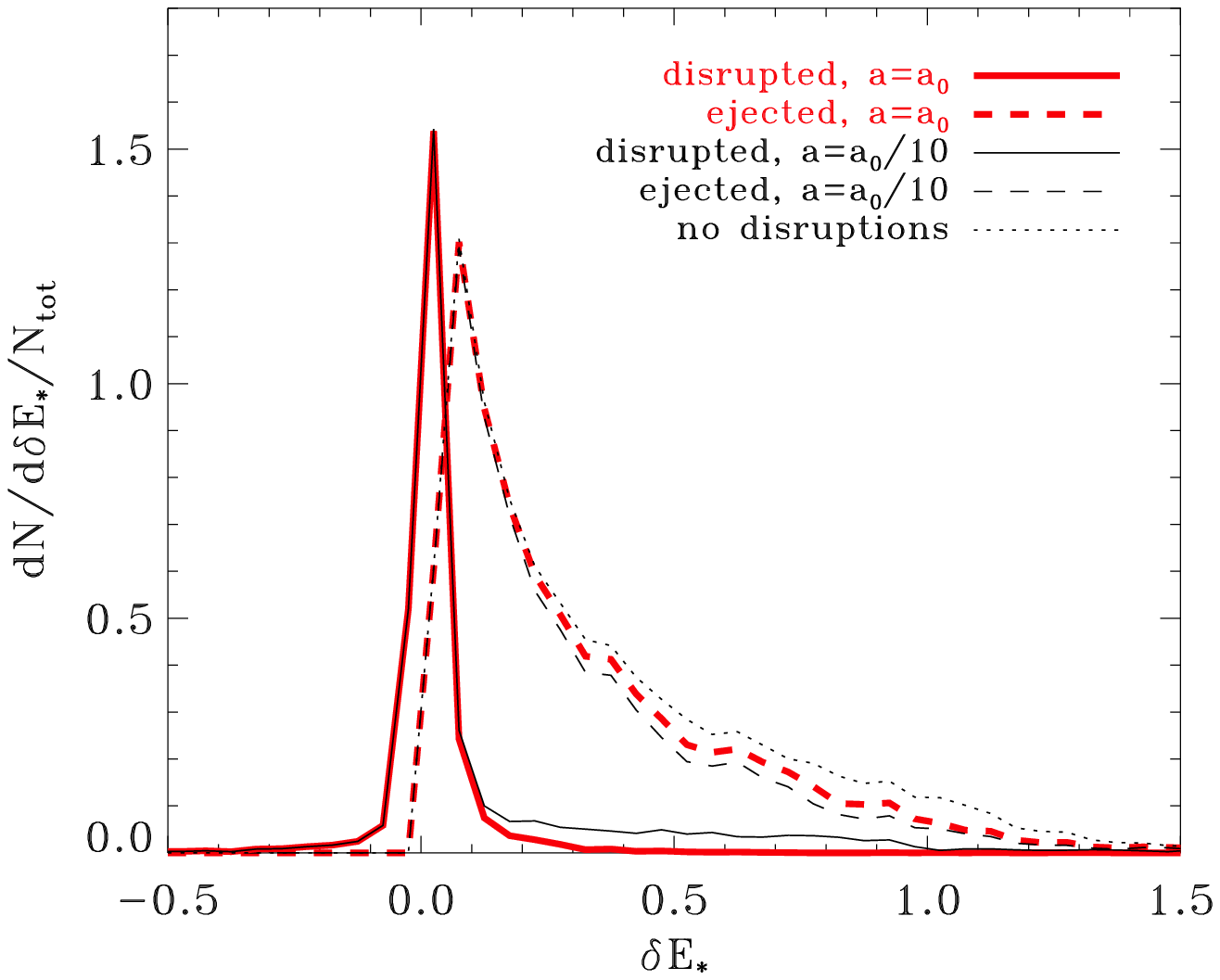}{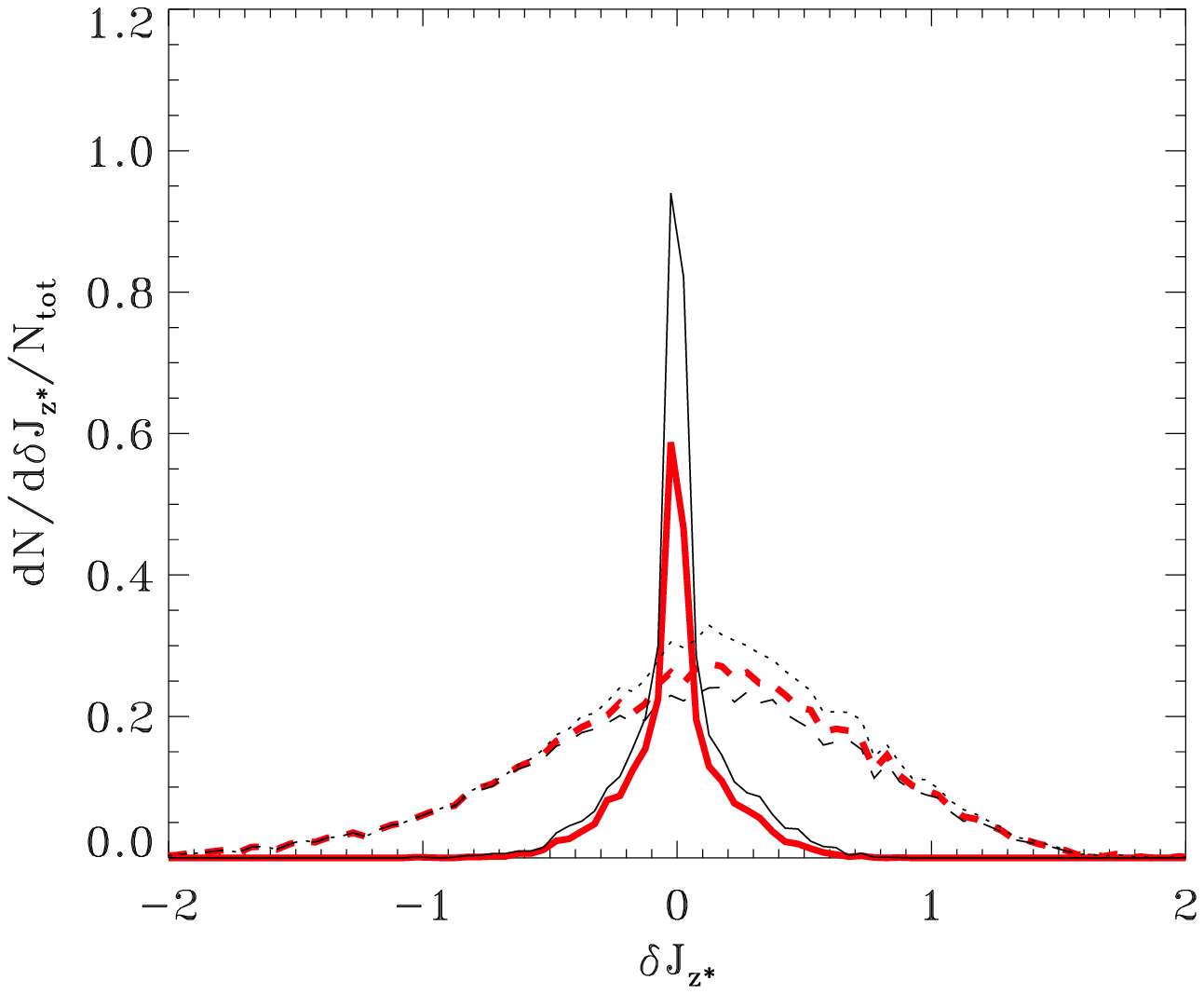}\\
\caption{Normalized distributions of changes in specific energy ({\it left panel}) and in the $z$-component of the specific orbital angular 
momentum ({\it right panel}) for the ejected ({\it dashed curve}) and disrupted ({\it solid curve}) stars when $e=0.1$. Energy is given in unit of 
$GM_{12}/a$, angular momentum in unit of $(GM_{12}a)^{1/2}$. Line styles as in Fig.~\ref{frac}.}\label{deltej}
\end{figure*}

Figure~\ref{frac} compares the fraction of stars that are ejected from the system with those that are disrupted in scattering experiments with $q=1/81$. Tidal disruptions produce two interesting effects: (1) when $a_*\sim a$ a significant fraction of stars experience strong interactions 
with the secondary hole and cross the tidal radius of $M_1$ before being ejected; (ii) frequent tidal disruptions occur even when $a_*\ll 
a$, a regime where ejections are rare. These are stars that are driven into the tidal loss cone by the Kozai mechanism. Tidal 
disruptions have then the double effect of partially suppressing stellar ejections (especially when the binary eccentricity is large) 
and at the same time of extending inward the influence domain of the binary (the $a_*/a$ interval where the black hole pair can alter the stellar cusp). 
Figure~\ref{deltej} shows the distributions of changes in specific energy and angular momentum ($z$-component) for the stars that are ejected and for those 
that are disrupted. Such distributions are narrowly peaked around zero in the case of the disrupted population, while are much broader and skewed towards 
positive values for the ejected component. The evolution of the MBHB is then determined by stellar ejections, since on average the disrupted stars do 
not exchange energy and orbital angular momentum with the binary.

\subsection{Hybrid model}
We finally describe our hybrid model. Given the binary-cusp system parameters $q$, $\gamma$, $M_7$, and $\sigma_{100}$, we calculate the 
corresponding value of the initial binary semimajor axis $a_0$. We use $a_*$ to describe the {\it absolute} semimajor axis of interacting
stars, and only consider the relevant portion of the cusp enclosed in the $a_*$ interval $[10^{-3}a_0,100a_0]$. This range is binned 
in 100 equally log-spaced bins labelled by the index $i$, and the initial stellar mass in each bin is given by $m_{i}=m_*\Delta a_{*,i}dn_0(a_{*,i})/da_*$ ($i=1,2,3,...,100$), where $a_{*,i}$ is the centroid of the $i$th bin and $\Delta a_{*,i}$ is the bin width. 
At $t=0$ the MBHB is at separation $a_0$ with eccentricity $e_0$ and period $P_0$; we then evolve the system numerically forward 
in time according to the equations
\begin{eqnarray}
a_{k+1}&=&a_k-\frac{\Delta \E_k}{\E_k}a_k,\label{aj}\\
e_{k+1}&=&e_k-\frac{1-e_k^2}{2e_k}\left({\Delta \E_k \over \E_k}+{2\Delta \J_k\over \J_k}\right),\label{ej}
\end{eqnarray}
where the index $k$ ($k\ge0$) labels the timestep, $a_k$ and $e_k$ are the binary semimajor axis and eccentricity, $E_k$ and $\J_k$ are 
the energy and angular momentum of the binary, and $\Delta$ refers to the variation in the $k$-th timestep $\Delta{t_k}$. The 
increments $\Delta\J_k$ and $\Delta\E_k$, depend on the mass $\sum_i\Delta m_{i,k}$ that interacts with the binary in the $k$-th 
timestep. The subtlety lies in properly extracting $\Delta m_{i,k}$ from the set of scattering experiments described in Section 3. 
Numerical experiments are carried at a fixed binary separation, and the relevant parameter in determining the fate of a star is the 
ratio $s=a_*/a$. In our runs we sample the range  $1/20<s<20$, and this interval is divided in equally log-spaced bins labelled by 
the index $j$ as $s_j$. For each $s_j$ bin, we construct the functions  $df/d\tau|_j$, $d{\mathcal E}/d\tau|_j$ and $d{\mathcal J}/d\tau|_j$, 
which are the differential fractions of ejected stars, mean energy exchange, and mean angular momentum exchange as a function of 
$\tau$, the time expressed in unit of the binary period. The trick is to assign to each bin $a_{*,i}$ the right $s_j$ value as the 
binary semimajor axis $a$ evolves, and to properly connect the physical time $t$ describing the evolution of the system to 
the `scattering experiment time' $\tau$ (expressed in units of $P$). For the time being, let us ignore, for simplicity, the eccentricity evolution. The integration scheme then proceeds as follows.

Consider the first timestep $\Delta{t_0}$. In each $a_{*,i}$ bin, the amount of ejected (or disrupted) mass{\footnote{Here 
we do not distinguish between ejected and disrupted stars. The distribution $df/d\tau$ is actually split in 
$df/d\tau_{\rm ej}$ and $df/d\tau_{\rm dis}$ to account for both components.}}~ in this first timestep is 
\begin{equation}\label{m0}
\Delta m_{i,0}=m_i\left[\frac{df}{d{\tau}}\bigg|_{j{_0}}(\tau=0)\,\frac{\Delta t_0}{P_0}\right],
\end{equation}
where $j_0$ identifies the $s_j$ bin to which the $a_{*,i}$ stellar bin belongs in the first timestep. If a particular 
$a_{*,i}$ bin lies outside the $1/20<s<20$ range, then it does not contribute in the evolution of the binary at the considered 
timestep. The binary separation $a$ is then evolved according to equation (\ref{aj}), where
\begin{equation}\label{E0}
\Delta{E_0}=\sum_i\left[\frac{d{\mathcal E}}{d{\tau}}\bigg|_{j{_0}}(\tau=0)\,\frac{\Delta t_0}{P_0}\right]\Delta m_{i,0}
\end{equation}
is given by summing over all $a_{*,i}$. We accordingly shrink the binary from $a_0$ to $a_1$.

Consider now the second timestep $\Delta t_1$. Since the stellar distribution changes with time, in
principle one should carry out new scattering experiments
according to the updated stellar distribution to derive $df/d\tau|_j$ at every timestep. However, as long as the stars depleted during the previous steps are appropriately excluded, the original scattering experiments can still be used. For the stars in a $s_j$ bin, the elapsed scattering-experiment time $\tau_{j,1}$ during the first timestep can be solved from the implicit equation 
\begin{equation}\label{1}
m_i\int_0^{\tau_{j,1}}\frac{df}{d{\tau}}\bigg|_{j{_1}}d{\tau}=\Delta m_{i,0},
\end{equation}
where $df/d\tau|_j$ is the same function as in the first timestep, and $j_1$ identifies the new $s_j$ bin to which the $a_{*,i}$ stellar bin belongs in the second timestep. In the second timestep, the time zero point of the function $df/d\tau|_{j_1}$ shifts to $\tau=\tau_{j,1}$ to exclude the stars with depletion timescales shorter than $\tau_{j,1}$, so the interaction mass becomes
\begin{equation}\label{m1}
\Delta m_{i,1}=m_i\left[\frac{df}{d\tau}\bigg|_{j{_1}}(\tau=\tau_{j,1})\,\frac{\Delta t_1}{P_0}\left(\frac{a_1}{a_0}\right)^{-3/2}\right],
\end{equation}
where $(a_1/a_0)^{-3/2}$ accounts for the change in the period of the binary as it shrinks from $a_0$ to $a_1$.
We then evolve again the binary according to equation (\ref{aj}), where now $\Delta{E_1}$ is given by 
\begin{equation}\label{E1}
\Delta{E_1}=\sum_i\left[\frac{d{\mathcal E}}{d{\tau}}\bigg|_{j{_1}}(\tau=\tau_{j,1})\frac{\Delta t_1}{P_0}
\left(\frac{a_1}{a_0}\right)^{-3/2}\right]\Delta m_{i,1}.
\end{equation}
For a generic timestep $\Delta t_k$, the interacting mass is
\begin{equation}\label{mk}
\Delta m_{i,k}=m_i\left[\frac{df}{d\tau}\bigg|_{j{_k}}(\tau=\tau_{j,k})\frac{\Delta t_k}{P_0}\left(\frac{a_k}{a_0}\right)^{-3/2}\right], 
\end{equation} 
where $j_k$ identifies the $s_j$ bin to which the $a_{*,i}$ stellar bin belongs in the $k$-th timestep, and $\tau_{j,k}$ 
labels the value of $\tau$ that solves the implicit equation:
\begin{equation}\label{1}
m_i\int_0^{\tau_{j,k}}\frac{df}{d{\tau}}\bigg|_{j{_k}}d{\tau}=\sum_{k'=0}^{k-1}\Delta m_{i,k'}.
\end{equation}
The binary is then evolved according to equation (\ref{aj}), where $\Delta{E_k}$ is given by replacing the index 1 with $k$ 
in equation (\ref{E1}). 

In this way we account for the fact that, in each stellar bin $a_{*,i}$, the interacting fraction in the timestep $k$ is governed 
by the stars left in the bin at that timestep following the ejection occurred in the previous timesteps; and that the ejection 
occurs at a rate given by the $s_j{_k}$ bin to which the $a_{*,i}$ stellar bin belongs at the $k$-th timestep. When we also 
consider the binary 
eccentricity evolution, we interpolate the values of $df/d\tau|_j$ and $d{\mathcal E}/d\tau|_j$ between the different eccentricities 
sampled by the scattering experiments $e=(0.1, 0.3, 0.6, 0.9$); the eccentricity is evolved according to equation (\ref{ej}), 
where $\Delta{J_k}$ is computed from the analogous of equation (\ref{E1}), but using $d{\mathcal{J}}/d\tau$.
We only considered the variation in the $z$-component of the angular momentum, because in a spherical stellar cluster the rotational Brownian motion of a MBHB 
is negligible \citep{merritt02a}.

Our hybrid approach relies on an adiabatic approximation, i.e. the MBHB orbital evolution is assumed to be slower compared to the typical star-binary 
interaction timescale. This is certainly true for chaotic encounters, but it is not so for secularly evolving stars. To justify our evolution scheme, 
we have therefore run test scattering experiments in which the MBHB was evolved by hand, according to the shrinking rates derived with the hybrid model.
The initial conditions of the test experiments were the same as in \S~\ref{properties},  and the stellar-disruption rates were calculated following 
the scheme described in \S~\ref{lcfr}. This setup allowed us to directly measure the disruption rate caused by an inspiralling binary on a population of 
stars drawn from a chosen density distribution. Figure~\ref{dr81} compares the stellar-disruption rates derived in this fashion to those given yielded by 
the hybrid model; the agreement between the two is quite good, validating our orbital integration scheme. 

\begin{figure}
\plotone{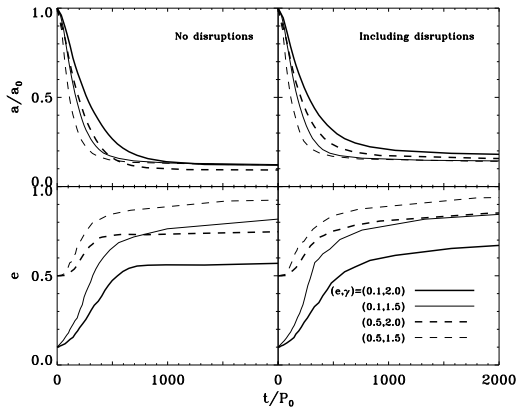}\\
\caption{Evolution of binary separation $a$ ({\it top panels}) and eccentricity $e$ ({\it bottom panels}) for a MBHB with different combinations of the 
initial $e,\gamma$ values. The curves in the left panels are computed neglecting tidal disruptions, while those in the right panels include 
the effect of stellar disruptions. The other system parameters are $q=1/81$, $M_7=1$, and $\sigma_{100}=1$.}\label{ae}
\end{figure}

\subsection{Results}

Figure~\ref{ae} shows the evolution of a MBHB with $q=1/81$, $M_7=1$, and $\sigma_{100}=1$ according to our hybrid model. The unit of time, $P_0$, is the 
initial binary orbital period at $a=a_0$, equal to $(400,6700)$ yr when $\gamma=(2,1.5)$. When stellar disruptions are not taken into account, the
orbital semimajor axis shrinks by a factor of $10$ on a timescale of $500P_0$ before the binary stalls: at the same time, the eccentricity 
increases significantly to $e\simeq0.5-1$, depending on the initial value and of the parameter $\gamma$. When compared with the central panel of Figure 7 
in \citet{sesana08}, the results of the two integration schemes appear to be in excellent agreement. The inclusion of stellar disruptions 
causes the binary to stall at slightly larger $a$ and higher $e$. The increase in the stalling radius is caused by the partial suppression of energetic 
ejections in favor of tidal disruptions. The larger eccentricity increase can be explained as follow. \citet{sesana08} demonstrated that stars with $a_*<a$ 
tend to drive the binary toward circularization, while stars with $a_*>a$ tend to increase its eccentricity. Since the former are the most susceptible to tidal 
disruptions, and disrupted stars do not exchange energy and angular momentum with the binary on the average, the relative contribution of stars with 
$a_*>a$ is larger in the presence of tidal disruptions, pushing the binary eccentricity to higher values. In a realistic situation, the shrinking of 
the binary would continue at $t>500P_0$ because of loss-cone diffusion processes and gravitational wave emission, which are not considered in this study.

\begin{figure}
\plotone{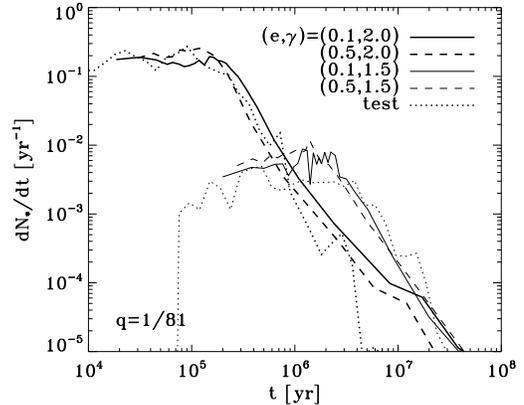}\\
\caption{Stellar disruption rates as a function of time for an evolving MBHB with  $q=1/81$, $M_7=1$, $\sigma_{100}=1$, and 
different combinations of the initial $e,\gamma$ values of the system. Each dotted curve is derived from $10^4$ test scattering experiments 
in which the MBHB is evolved according to the rates given by the hybrid model (see the solid lines in the right panels of  Fig.~\ref{ae}). 
Other line styles as in Fig.~\ref{ae}}.
\label{dr81}
\end{figure}

Figure~\ref{dr81} shows the stellar-disruption rates predicted by the hybrid model (solid and dashed lines) together with those derived by the test scattering 
experiments with the MBHB evolved by hand (dotted lines). During the first $500P_0$, the rate remains constant,
at a level that is comparable to the peak value calculated for a stationary binary. The duration of the plateau, however, is longer 
in the case of a decaying pair, as new stars are continuously added to the time-varying interaction loss cone. At $t\ga 500P_0$, the evolution time 
of the MBHB exceeds the tidal disruption timescale, strongly-interacting stars get depleted, and  the consumption rate drops sharply.
The figure also shows that the peak disruption rate is not sensitive to the binary eccentricity but depends on $\gamma$ according 
to the scaling relation in equation~(\ref{ndotscale}).
After $10^8$ yr, the total number of disrupted stars is $(6.5\times10^4,2.3\times10^4)$ for $\gamma=(2,1.5)$. The disruption rate during
the plateau phase remains constant even though equation (\ref{ndotscale}) predicts an increase $\propto a_0/a$ (assuming an isothermal 
cusp). Equation (\ref{ndotscale}) was derived, however, assuming a stationary binary at different orbital separations in an unperturbed stellar 
profile. In our hybrid model, the binary shrinks while depleting the stellar cusp, and many of the stars available for disruption at, say, $a=0.1a_0$ 
are actually ejected or disrupted during the evolution of the pair from from $a_0$ to $0.1a_0$, leveling off the disruption rate. 
Given that: (1) the disruption rate in the plateau phase remains constant and is consistent with the predictions of the Kozai mechanism even
for evolving binaries; and (2) the duration of the plateau is of the order of the binary decay timescale,
$t_d\propto q^{-3/2}P_0$ \citep{sesana08}, the total number of disrupted stars can be scaled according to 
\begin{eqnarray}\label{ntotscale}
N_* \propto t_d \dot{N}_* &\propto& (3-\gamma)^{-1/2}q^{(2-\gamma)/(6-2\gamma)}M_1^{2/3}\sigma_* \nonumber\\
&\propto& (3-\gamma)^{-1/2}q^{(2-\gamma)/(6-2\gamma)}M_1^{11/12}, 
\end{eqnarray}
where we used equation (\ref{period}) for $P_0$ and the $M_1-\sigma_*$ relation in the second proportionality. According to the above 
equation, for $q=1/81$, $N_*$ should drop by a factor of $2.5$ as $\gamma$ varies from $2$ to $1.5$, consistent with the numbers 
derived from our hybrid model. Also, as long as $\gamma\ga 1.5$, $N_*$ should not be very sensitive to the binary mass ratio $q$. 
We stress that these scalings are derived from the no-GR/no-cusp experiments, and their validity is limited to binaries with $q>0.01$.

\section{Summary and conclusions}\label{dc}

In this paper, we have studied the tidal disruption rate in a system composed by a MBHB and a bound stellar cusp.  We have carried 
out numerical scattering experiments for a detailed investigation of the mechanisms responsible for the repopulation of the tidal loss cone, and
developed a hybrid model to self-consistently solve for the evolutions of the binary, the depletion of the stellar cusp, and the 
stellar consumption rate. Our main results can be summarized as follows:
\begin{enumerate}

\item For unequal binaries ($q < 0.1$), the tidal disruption cross section for bound stars, which quantifies the probability of stellar 
disruption, is three orders of magnitude larger than the cross section for a single MBH fed by two-body relation. Two mechanisms contribute to 
such enhancement, the Kozai secular effect and chaotic resonant interactions. When the eccentricity of the MBHB is small, stars inside 
the Kozai wedge repopulate the tidal loss cone on the Kozai timescale, while stars outside the Kozai wedge but inside the interaction loss 
cone are scattered into the tidal loss cone at random times due to close interactions with the secondary hole. When the eccentricity is 
large, chaotic loss-cone repopulation becomes dominant over the entire range of stellar semimajor axis $a_*\ga (1-e)a$.

\item GR and cusp-induced precession quench the Kozai secular evolution of interacting stars, causing a significant suppression (by a factor of $\sim 10$) 
of the disruption rate for $q < 0.01$. Therefore, the optimal enhancement of the tidal disruption rate by a MBHB  occurs for mass ratios $0.01 < q < 0.1$. 
Note that even if suppressed by a factor of $\sim 10$, the tidal disruption rate for binaries with $q < 0.01$ is still two order of magnitude larger 
than that given by standard relaxation processes around a single MBH. 

\item If a MBHB with mass ratio $q\ll1$ does not evolve significantly during $1/q$ revolutions, tidal disruptions of bound stars could 
initially persist at a constant rate (``plateau phase") that is four dex higher than the typical rates predicted for single MBHs. After one Kozai timescale 
(evaluated at $a_*=a$), the tidal loss cone is repopulated mainly by chaotic interaction, and the stellar disruption rate 
decreases with time. The majority of stars are disrupted during a post-plateau later phase.

\item If a MBHB evolves significantly on a timescale of $1/q$ revolution, the plateau phase of stellar disruptions may last longer than 
a Kozai timescale. Tidal disruptions of bound stars slow down the shrinking of the binary and speed up the growth of binary eccentricity.

\end{enumerate}

Our results indicate that, after the formation of an unequal-mass MBHB at the center of a dense stellar cusp, the tidal disruption rate may 
go through three distinct evolutionary phases. The first phase begins shortly after the MBHs become bound, and is characterized by a 
disruption rate as high as $0.1-1$ stars per year, resulting from the three-body interactions between the binary and the bound stars \citep{chen09}. 
When the decay timescale of the MBHB becomes longer than the tidal disruption timescales of stars with $a_*\sim a$, a second phase starts, 
where cusp depletion from slingshot ejections and tidal disruptions causes a sharp drop in the disruption rate. \citet{chen08} showed that, 
unless stellar relaxation is far more efficient than two-body ``collisions", the tidal disruption rate in this phase is orders of magnitudes 
lower than typical for a single MBHs. A third phase begins if the MBHB shrinks to the gravitational wave regime and eventually coalesces. 
The tidal disruption rate then gradually increases to the value typical for single MBHs, $10^{-5}-10^{-4}\,\ndot$, within one stellar 
relaxation timescale \citep{merritt05}. The number of stars disrupted during phase I is about $10^4-10^5$ for $M_7=1$ and $q=1/81$. The number of stars
disrupted in phases II and III depends on the efficiency of stellar relaxation, but would not significantly exceed $\sim10^5-10^6$. 
If a galaxy formed, on the average, one unequal-mass MBHB following a minor merger in its lifetime, then the above numbers imply that in a 
sample of tidal flares from MBHs of $\sim10^7\msun$, about $10\%$ of events would be associated to binaries. If a galaxy forms unequal-mass 
MBHBs multiple times during its lifetime, then the detection rate of tidal events from binaries in transient surveys may be higher. 
Given the very high rates, there is also the possibility to identify MBHBs in galaxies hosting multiple tidal flares within a years-to-decades 
time span. Over the next decade, synoptic surveys are expected to detect hundreds of tidal disruption candidates.  A tidal flare associated to a MBHB is likely interrupted within one orbital period of the binary \citep{liu09}, therefore is distinguishable from the flares produced by single MBHs, as long as the orbital period of the binary is shorter than the duration of a transient survey. If MBHB-driven disruptions 
account for $10\%$ of the total rate, then the prospects of identifying MBHBs through tidal flares are promising. Because the predicted rates in the 
three phases are significantly different from one another, the average stellar disruption rate over the lifetime of a galaxy is sensitive to the infalling rate of secondary MBHs and the relative duration of each phase. A comparison between the observational detection rate of tidal 
events \citep{donley02,gezari08} and those predicted during the three phases may then shed light on the abundance and dynamical evolution of MBHBs. 

\acknowledgments
Support for this work was provided by NASA through grant NNX08AV68G (P.M.). X.C. and F.K.L. thank the Chinese national 973 program (2007CB815405), the Research Fund for the Doctoral 
Program of Higher Education (RFDP), and the China Scholarship Council for financial support. We also acknowledge support from the National Natural Science Foundation of China (11073002). 
We are grateful to J. Magorrian and F. Haardt for early discussions on this topic. We also thank the referee whose suggestions really contributed to improve the quality of the manuscript. The scattering experiments were performed on the SGI Altix 330 system at the Astronomy 
Department, Peking University and the Pleiades cluster at the Department of Astronomy \& Astrophysics, University of California, Santa Cruz.

\end{document}